\documentclass[twocolumn,showpacs,preprintnumbers,amsmath,amssymb]{revtex4}
\usepackage[dvips]{graphics}
\usepackage{color}

\newcommand{\beq}{\begin{equation}}
\newcommand{\eeq}{\end{equation}}
\newcommand{\bma}{\begin{math}}
\newcommand{\ema}{\end{math}}
\newcommand{\beqa}{\begin{eqnarray}}
\newcommand{\eeqa}{\end{eqnarray}}

\def\expect#1{\langle\, #1\, \rangle}

\def\opone{\le\textbf{}\textbf{}avevmode\hbox{\small1\kern-3.8pt\normalsize1}}

\newcommand{\be}[1]{     \begin{eqnarray} \mbox{$\label{#1}$}   }

\newcommand{\ee}{\end{eqnarray}}
\newcommand{\pref}[1]{(\ref{#1})}

\newcommand\grad{\vec\nabla}
\newcommand\half{\frac 1 2 }

\newcommand\ket [1] {|#1 \rangle }

\begin{document}

\title{Quantum Hall system in Tao-Thouless limit}

\author{E.J. Bergholtz}

\author{A. Karlhede}

\affiliation{Department of Physics,
Stockholm University \\ AlbaNova University Center\\ SE-106 91 Stockholm,
Sweden}

\date{\today}

\begin{abstract}
We consider spin-polarized electrons
in a single Landau level on a torus.  The quantum Hall
problem is mapped onto a one-dimensional lattice model with lattice constant
$2\pi/L_1$, where $L_1$ is a circumference of the torus (in
units of the magnetic length). In the Tao-Thouless limit,  $L_1\rightarrow 0$, the
interacting many-electron problem is exactly diagonalized at any
rational filling factor $\nu=p/q\le 1$.  For odd $q$, the ground state has the
same qualitative properties as a bulk ($L_1 \rightarrow \infty$)
quantum Hall hierarchy state and the lowest energy quasiparticle
exitations have the same fractional charges as in the bulk. These
states are the $L_1 \rightarrow 0$ limits of the Laughlin/Jain wave
functions for filling fractions where these exist. We argue that the
exact solutions generically, for odd $q$, are continuously 
connected to the two-dimensional bulk quantum Hall hierarchy states, {\it ie}
that there is no phase transition as $L_1 \rightarrow \infty$ for filling factors where 
such states can be observed. For
even denominator fractions, a phase transition
occurs as $L_1$ increases. For $\nu=1/2$ this leads to the system
being mapped onto a Luttinger liquid of neutral particles at small
but finite $L_1$, this then develops continuously into the
composite fermion wave function that is believed to describe the
bulk $\nu=1/2$ system. The analysis generalizes to non-abelian
quantum Hall states.

\end{abstract}
\pacs{73.43.Cd, 71.10.Pm, 75.10.Pq}

\maketitle

\section{Introduction}\label{introduction}

The two-dimensional electron gas in a perpendicular magnetic field,
the quantum Hall (QH) system, is remarkably rich. The integer QH
effect shows a conductance precisely quantized in integer units of
$e^2/h$ \cite{vonk}, and in the fractional QH regime the  conductance is
quantized in fractions of $e^2/h$ and there are fractionally charged
quasiparticles that obey (abelian) fractional statistics \cite{tsui,stat1, stat2}. Whereas
the integer effect can be understood in terms of non-interacting
electrons \cite{it1,it2}, the fractional effect is caused by the interaction
between the electrons \cite{Laughlin83}. Other states observed in this
strongly correlated electron system are metallic states that
resemble a free two-dimensional Fermi gas \cite{jiang,willett90,wang} and inhomogenous striped
states in higher Landau levels \cite{striped1, striped2}.  A further example of an exotic
quantum state that may form is a non-abelian state
where the quasiparticles obey non-abelian fractional statistics
\cite{mr,nayak,readrezayi}.

The fractional QH effect is understood as an incompressible quantum liquid with fractionally charged quasiparticles; this is
 based on Laughlin's wave functions for filling fractions  $\nu=1/(2m+1)$ \cite{Laughlin83}. For other fractions,
a hierarchy construction where quasiparticles condense to form new
quantum liquids just as electrons form the Laughlin states was
proposed by Haldane \cite{hierarchyHaldane},  Laughlin \cite{hierarchyLaughlin}   and by Halperin
\cite{hierarchyHalperin}. An alternative view, where electrons
supposedly capture magnetic flux to form composite fermions that see
a reduced magnetic field was developed by Jain \cite{jain89,jainrev1,jainrev2,jainbook}. Successful
mean field theories that support this idea of flux attachment have
been developed \cite{zhang,read89, lopez}. The composite fermion approach has
received strong support by experiments performed in the half-filled Landau level.
Near $\nu=1/2$, ballistic transport is consistent with
particles moving in a reduced magnetic field in accordance with the
composite fermion prediction \cite{ballistic1,ballistic2,ballistic3}.\footnote{However, an
alternative consistent interpretation is that it is a quasiparticle,
with a reduced charge compared to the electron, that moves in the
original full magnetic field; what is measured is the product of the
charge and the magnetic field.} At $\nu=1/2$, the external magnetic
field is completely absorbed by the electrons and the composite
fermions see no magnetic field---they form a two-dimensional Fermi
gas. The mean field theory of this state \cite{hlr,kalmeyer} is in excellent agreement
with surface acoustic wave experiments
\cite{willett90, Willettreview}.

In spite of these impressive results it is our opinion that there
are basic questions concerning the QH system that remain to be
answered.\footnote{A critical discussion of the status of the theory
of the QH effect is presented by Dyakonov \cite{Dyakonov}.} Most
importantly, a microscopic understanding is lacking. It is true that
good many-body wave functions exist, but there is no microscopic
understanding or derivation of them. For example, according to the
composite fermion picture the fractional QH ground state is formed
when composite fermions fill an integer number of  effective Landau
levels in the reduced magnetic field.  No derivation of this
scenario from the many body wave function exists. Mean field theory,
which is successful,  does give support to the idea of electrons
binding flux quanta; however, it is not understood why mean field
theory works as well as it does. A further question is the relation between
the original hierarchy description and composite fermions; are they
alternative descriptions of the same thing as argued by Read
\cite{read90} and by Blok and Wen \cite{blokwen}, or are they fundamentally different as
argued by Jain \cite{jain89,jainrev1,jainrev2,jainbook}? Our aim is to contribute to a solution to
these problems.

In this article we consider spin-polarized interacting electrons
within a single Landau level. It has recently been realized that
there is a limit in which this problem can be exactly solved for any
rational filling factor $\nu=p/q\le1$---and that the solution is
physically relevant \cite{bk1,bk2,we06,hierarchy,conformal}.
We here expand on, and provide details of, our work presented in Refs. \onlinecite{bk1,bk2}.
We consider the interacting electron
gas on a torus where it  becomes equivalent to a one-dimensional
lattice model with a complicated long-range interaction. The
solvable limit is the thin torus: $L_1\rightarrow 0$, where $L_1$ is
one of the circumferences of the torus (the other circumference
being infinite). In this limit  the interaction in the
one-dimensional lattice problem becomes purely electrostatic and the
ground state is a "crystal" of electrons occupying fixed positions
on the lattice as far apart from each other as possible. For
$\nu=1/3$ every third site is occupied; interestingly enough this is
the state introduced by Tao and Thouless in 1983 to explain the
fractional QH effect \cite{Tao83} and we call these crystal states Tao-Thouless
(TT) states. The fractionally charged quasiparticles are domain walls
separating the degenerate ground states that exist on the torus.

It should be noted that the mapping of the lowest Landau level, or of any single Landau
level for that matter, onto a one dimensional lattice problem is exact and that varying $L_1$ may
alternatively be thought of as varying a parameter in the hamiltonian that controls the range
of the interaction while keeping the two-dimensional space fixed (and possibly infinite). To stress this
we refer to the thin limit as the Tao-Thouless (TT) limit. If the ground state
in the TT-limit remains the ground state as $L_1 \rightarrow \infty$ we may thus conclude that the
experimentally accessible ground state is adiabatically connected to the ground state in the TT-limit.

The simple limit $L_1\rightarrow 0$ may at first seem to be of little physical interest---it is really an extreme case:
the interaction is both very short range and anisotropic and is furthermore purely electrostatic.
The surprising fact is that the ground state in this limit has all the qualitative properties of a
fractional quantum Hall state, such as a gap, the correct quantum numbers and quasiparticles with
the correct fractional charge. We argue that the simple
TT-ground states obtained in the TT-limit develop continuously, without a phase transition,
into the fractional QH hierarchy states, as $L_1 \rightarrow \infty$, for filling factors where such states are observed.
Thus, we argue that the TT-state in general describes the fractional QH phase observed as
$L_1\rightarrow \infty$, in the sense that these states are adiabatically connected.
We show that the TT-states are the $L_1 \rightarrow 0$ limits of the
Laughlin and Jain wave functions for filling fractions where these exist. In the TT-limit the original hierarchy construction is manifest: the TT-ground states are condensates of quasiparticles.

The hierarchy structure of states in the TT-limit has recently emerged within
a conformal field theory (CFT) construction of bulk hierarchy wave functions for all fractions that
are obtained by successive condensation of quasielectrons (as opposed to quasiholes) \cite{hierarchy,conformal}.
These wave functions reduce to the correct TT-states in the TT-limit and are obtained by a
natural generalization of the conformal construction of the composite fermion wave functions \cite{hans}.
This supports the adiabatic continuity fro
m the TT-limit to the bulk for general hierarchy states.

The TT-states are the ground states also for the even denominator fractions in the TT-limit;
however, for these fractions  we claim that there is always a phase transition when $L_1$ increases.
This is supported by numerical studies and by a detailed analysis of $\nu=1/2$, which we believe is
a representative case for the even denominator fractions.

At $\nu =1/2$, there is a phase transition from the gapped TT-state
to a one-dimensional gapless state at $L_1 \sim 5$. (Lengths are
measured in units of the magnetic length, $\ell= \sqrt{\hbar c/eB}=1$.) Also for
this gapless phase there is an exact solution:  For a hamiltonian
that is a good approximation at $L_1\sim 5$, the low energy sector
consists of non-interacting neutral fermions (dipoles); the ground
state is a one-dimensional Fermi sea and there are gapless neutral
excitations. This provides an explicit example of
interacting electrons in a magnetic field being equivalent to free
particles that do not couple to the magnetic field. The ground state in
the exact solution is a version of the composite fermion state
\cite{jainbook} given by Rezayi and Read \cite{rr}. There is strong
numerical evidence \cite{bk2} that this gapless one-dimensional
state develops continuously, without a phase transition, into the
two-dimensional bulk version of the Rezayi-Read state that is
believed to describe the observed metallic state at $\nu =1/2$.

The Moore-Read pfaffian state \cite{mr}, believed to describe the half-filled second Landau level, $\nu=5/2$, also exists in the TT-limit.
A simple construction gives the quarter-charged quasiholes and quasielectrons as domain walls between the six-fold degenerate
ground states and the non-trivial degeneracies for these excitations is obtained \cite{haldaneAPS,we06}.
Similar results for other non-abelian states exist \cite{seidel06, read06, weunpublished}; see also Ref. \onlinecite{haldanejack}.

Over the years there have been many interesting attempts to improve
the understanding of the QH system and it is impossible to here
mention them all.  In our work what emerges is a one-dimensional
theory of the quantum Hall system which depends on a dimensionless
parameter $L_1$, where the Tao-Thouless states \cite{Tao83} are the
exact solutions as $L_1 \rightarrow 0$ and the fractionally charged
quasiparticles are domain walls between the degenerate ground
states.  The TT-states are adiabatically connected to bulk QH states
as $L_1 \rightarrow \infty$---there is no phase transition if a QH
state is observed. Searching the literature one finds hints and
suggestions for such a scenario. Anderson noted in 1983 that
Laughlin's wave function has a broken discrete symmetry and that the
quasiparticles are domain walls between the degenerate ground
states \cite{anderson}. Furthermore he noted that the TT-state is
non-orthogonal to the Laughlin state and suggested that it can be
thought of as a parent state that develops into the Laughlin wave
function, without a phase transition, as the electron-electron
interaction is turned on. In 1984, Su concluded, based on exact
diagonalization of small systems on the torus, that the QH state at
$\nu=p/q$ is $q$-fold degenerate and that the lowest energy
excitations are quasiparticles with charge $\pm e/q$ that are domain
walls between the degenerate ground states \cite{su1984,su2}. In
1994, Rezayi and Haldane studied the Laughlin wave function on a
cylinder as a function of its radius and noted that it approaches
the Tao-Thouless state on the thin cylinder.  Implicit in their work
is the fact that the Laughlin wave function is the exact and unique
ground state to a short range pseudo-potential interaction on a
cylinder for any circumference \cite{Haldane94}. In retrospect, this
makes a very strong case for an adiabatic evolution from the
TT-state to the Laughlin wave function. A one-dimensional approach
to QH states was also considered by Chui in 1985
\cite{chui1985,chui86}. More recently, this has been explored in
connection with Bose-Hubbard models by Heiselberg \cite{heiselberg},
and it should also be mentioned that Dyakonov presents a
one-dimensional toy QH-model \cite{Dyakonov}.

At $\nu=1/2$, we find that the low energy sector consists of weakly
interacting dipoles. This relates to earlier descriptions in terms
of dipoles, in particular to the field theory of Murthy and Shankar
\cite{ms} and the work by Read
\cite{readearly,read.5}, Pasquier and Haldane \cite{pasq}, Lee \cite{dhlee} and Stern {\it et al} \cite{stern};
for reviews see Ref. \onlinecite{rev1}.
It is of course also reminiscent of composite fermions in general
\cite{jainbook} in that the particles do not couple to the magnetic
field and are weakly interacting.

We would also like to draw the attention to the construction of composite fermion  wave functions
directly in the lowest Landau level by Ginocchio and Haxton \cite{haxton}, 
the series of work by Wojs, Quinn and collaborators, see Ref.  \onlinecite{wojs} and references therein.
Explicit wave functions have been obtained within the original hierarchy construction by Greiter  \cite{greiter94}. 

The content of the article is as follows. The one-dimensional lattice model for 
interacting electrons in a single Landau level is introduced in Sec. \ref{model}.
In Sec. \ref{solution} we solve this problem exactly in the TT-limit
$L_1\rightarrow 0$, {\it ie} we diagonalize the interacting electron
hamiltonian for any rational filling factor in this limit. The
ground states and the fractionally charged quasiparticles are
identified and it is found that the former are condensates of the
latter thus proving the original hierarchy construction in this
limit \cite{hierarchy}. The quasiparticles are domain walls between the degenerate
ground states and their charge is determined by the Su-Schrieffer
counting argument. The energy of a quasielectron-quasihole pair is
determined as $L_1\rightarrow 0$ and it is found that the gap to
creating an infinitely separated such pair at $\nu=p/q$ decreases
monotonously with increasing $q$ but is independent of $p$; this is
in surprisingly good agreement with experiments, which are performed
in the two-dimensional bulk system, $L_1 \rightarrow \infty$. The
Laughlin and Jain fractions, as well as those observed by Pan {\it et al}
\cite{pan} are considered explicitly as examples, and finally the
relation to composite fermions and emergent Landau levels is
commented on.

The transition to the two-dimensional bulk system as $L_1
\rightarrow \infty$ is considered in Sec. \ref{bulk}, first for odd
denominator fractions leading to the QH hierarchy states, then for
the half-filled Landau level; comments on other even denominator
fractions and on non-abelian states are included. The conclusion is
that the rich structure and different phases of matter present in
the QH system exist also in the TT-limit, where it can be studied in
detail starting from a microscopic hamiltonian.

Technical details are burried in a series of appendices. The
mathematics that we need for a single Landau level on a torus
\cite{Haldane85PRL} including the construction of the lattice
hamiltonian is given in Appendix \ref{appA}. In Appendix \ref{appB}
we prove that the relaxation procedure given in Sec. \ref{solution}
actually gives the ground state. In Appendix \ref{appC}, we show
that the quasiparticle charge at $\nu=p/q$ is $e^*=\pm e/q$; the
result is obtained in the limit $L_1\rightarrow 0$ using the
Su-Schireffer counting argument. The energy of a
quasielectron-quasihole pair is obtained in Appendix \ref{appD}. In
Appendix \ref{appE} we show that the $L_1\rightarrow 0$ limit of
Laughlin's and Jain's wave functions are the TT-ground states
obtained in Sec. \ref{solution}. Appendix \ref{appF} shows that
Laughlin's wave function at $\nu=1/(2m+1)$ is the exact ground
state, and that there is a gap to excitations, for all $L_1$ for a
short-range interaction. Appendix \ref{appG} contains details of the
exact solution at $\nu=1/2$.

\section{Model}\label{model}

\begin{figure}[h!]
\begin{center}
\resizebox{!}{28mm}{\includegraphics{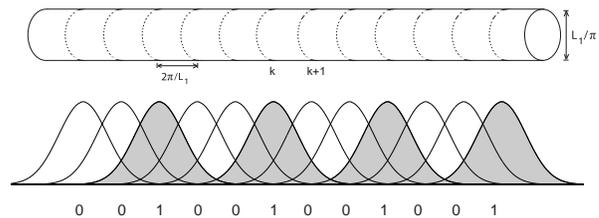}}
\end{center}\caption{\textit{{\small  Mapping of  a single Landau level onto a
one-dimensional lattice model. The two-dimensional space is a cylinder with
circumference $L_1$ (or a torus if the ends are identified) (top panel). The single particle wave functions are centered
along  circles and are gaussians of width one along the cylinder (central panel); they  are numbered by their momenta $k2\pi/L_1$
which also give the position along the cylinder. This
gives a one-dimensional lattice model with lattice constant $2\pi/L_1$ where each site is either empty (0)
or occupied by en electron (1). As an example the Tao-Thouless state 001 at  $\nu=1/3$ is shown (bottom panel).
}}}\label{mapping}
\end{figure}

We consider a single Landau level of spin-polarized electrons on a
torus with lengths $L_1,L_2$ in the $x$ and $y$-directions
respectively. We use units such that $\hbar= c/eB=1$ and choose
one-particle states $\psi _k$, $k=0,1,\dots, N_s-1$,  $N_s= L_1
L_2/2\pi$, that have $x$-momentum $2\pi k/L_1$ and are gaussians
centered at $y=-2\pi k/L_1$, see Appendix \ref{appA}. This provides
an exact mapping of the Landau level onto a one-dimensional lattice
model with lattice constant $2\pi/L_1$, see Fig. \ref{mapping}. Each
site can be either empty, 0, or occupied by an electron, 1. Thus a
basis of many particle states is provided by $\{| n_0 n_1 n_2\dots
n_{N_s-1} \rangle \}$, where $n_i=0,1$; alternatively,  states can
be characterized by the positions (or, equivalently, the
$x$-momenta) $\{k_1,k_2,...,k_{N_e}\}$ of the particles.

\begin{figure}[h!]
\begin{center}
\resizebox{!}{14mm}{\includegraphics{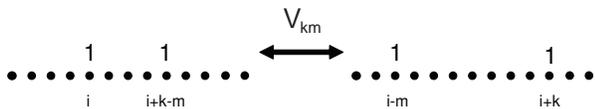}}
\end{center}\caption{\textit{{\small 
A general translationally
invariant two-electron interaction consists of the terms
where two electrons preserve the position of their center of mass. There are electrostatic terms $V_{k0}$, where the electrons do not move, and
hopping terms $V_{km}, \, m\neq 0$, where the electrons hop symmetrically.
}}}\label{hamilton}
\end{figure}

A general two-body interaction, $V(r)$,  that depends on the
distance, $r$, between two electrons only, such as a Coulomb or a
short-range delta function interaction, leads to the one-dimensional
hamiltonian
\be{ham} H&=&\!\!\sum_{n=0}^{N_s-1} \sum_{|m| <
k\leq N_s/2}\frac{V_{km}}{\!1\!+\!\delta_{k,\frac{N_s}2}\!\!}
c^\dagger_{n+m}c^\dagger_{n+k}c_{n+k+m}c_n \ .
\ee
Here $c^{\dagger}_k$ creates an electron in state $\psi_k$, $\{c_k,c_m^\dagger \}=\delta_{km}$, and $V_{km}=V_{k,-m}$.  $H$ consists of the
two-body terms that preserve the $x$-momentum, {\it ie}, the position of the center of mass of the electron pair, see Fig. \ref{hamilton}.
$V_{km}$ is the amplitude for two electrons separated $k-m$ lattice constants to hop symmetrically to a separation of
$k+m$ lattice constants.  $V_{k0}$ is the electrostatic repulsion (including the exchange interaction) between two
electrons separated $k$ lattice constants.\footnote{Note that $V_{N_s/2,m}$ is multiplied by $1/2$ because of the delta-function---this compensates for the fact that
for even $N_s$ every process with $k=N_s/2$ appears twice in (\ref{ham}).}

For a given real-space interaction $V(r)$, when $L_1$ is small the
lattice constant $2\pi/L_1$ is large and, hence, the dominant
$V_{km}$ are those with small $k,m$.
Furthermore, the wave functions $\psi_k$ are gaussians in the
$y$-direction, with width of the order of the magnetic length, hence
their overlap vanishes rapidly as $L_1 \rightarrow 0$, and the
electrostatic terms $V_{k0}$ dominate in
this limit. The physics is thus very
simple in the TT-limit $L_1 \rightarrow 0$---it is determined by
electrostatic repulsion only. In the two-dimensional limit,
$L_1\rightarrow \infty$, the range of the one-dimensional
interaction measured in number of lattice constants goes to infinity
for {\it any} real space two-dimensional interaction; this is true
also for a local interaction such as $V(r)\propto \bigtriangledown
^{2n} \delta ^2({\bf r})$.

The symmetry analysis of many-electron states on the torus was given by Haldane \cite{Haldane85PRL};  a simple version adapted to our needs can be found in
Appendix \ref{appA}. We give here the results. There are two translation operators, $T_\alpha$, $\alpha =1,2$,  that commute with the hamiltonian $H$; they
obey $T_1T_2=e^{2\pi i p/q}T_2T_1$, see (\ref{T1T2}).  These operators have
eigenvalues $e^{2\pi iK_{\alpha} /N_s}, K_{\alpha}=0,...,N_s-1$. $T_1$ corresponds to $x$-translations
and $K_1=\sum_{i=1}^{N_e} k_i$ is the total $x$-momentum (in units of $2\pi /L_1$).
$T_2$ translates the system one lattice constant in the $y$-direction, {\it ie} along the one-dimensional lattice and increases $K_1$ by $N_e$. At filling factor
$\nu=p/q$ (where $p$ and $q$ are relatively prime) $T_2^q$ commutes with $T_1$:  $\{H, T_1, T_2^q\}$ is a maximal set of commuting operators.
$T_2^k, k=0,1, \dots q-1$, generate $q$ degenerate orthogonal states, which have different $K_1$, when acting on any state---this is the $q$-fold center of mass degeneracy. Hence, each energy eigenstate is (at least) $q$-fold degenerate
and we choose to characterize it  by the smallest $K_1$. Thus, the energy eigenstates are characterized by a two-dimensional
vector $K_\alpha=0,...,N_s/q-1$, where $e^{2\pi i K_2 q/N_s}$ is the $T_2^q$-eigenvalue.

\section{Exact solution\label{solution}}

In this section we solve the problem of interacting spin-polarized electrons exactly at any rational filling factor $\nu=p/q\le1$ in the Tao-Thouless
limit; we diagonalize
the hamiltonian \pref{ham} and construct explicitly the ground state as well as the low energy charged excitations which turn out to have charge $\pm e/q$.
As noted above, the hopping elements,
$V_{km}, \, m\neq 0$,  vanish rapidly as $L_1$ decreases, whereas the electrostatic elements, $V_{k0}$, decrease much more slowly. 
Hence in the limit $L_1 \rightarrow 0$ effectively  only the electrostatic interaction survives and the hamiltonian \pref{ham} becomes
\be{elstat}
H_0=\sum_{i=0}^{N_s-1} \sum_{1\leq k \leq N_s/2} \!\frac{V_{k0}}{\!1\!+\!\delta_{k,\frac{N_s}2}\!\!}\  \hat n_i \hat n_{i+k} \ \ ,
\ee
where $\hat n_i=\hat c_i^\dagger \hat c_i$ and periodic boundary conditions have been imposed,  $\hat n_{i+N_s}=\hat n_{i}$.
$H_0$ in \pref{elstat} defines the Tao-Thouless limit.
At filling fraction $\nu=p/q=N_e/N_s$,
the energy eigenstates are simply the states
\be{states}
\{|n_0n_1\dots n_{N_s-1}\rangle \} \ \  , \ \ \ n_i=0,1 \ \ ,
\ee
where $N_e$ electrons occupy fixed positions on $N_s$ lattice sites,
\be{Hcondition}
 \sum_{i=0}^{N_s-1} n_i=N_e=N_sp/q \ \ .
\ee

\subsection{Ground states, quasiparticles and the hierarchy }\label{hierarchy}

Here we determine the ground state at $\nu=p/q\le1$  in the TT-limit,  {\it ie} for the
hamiltonian $H_0$ in \pref{elstat}.
This is the classical electrostatics problem of finding the position of $N_e$ electrons,
on a circle with $N_s$ sites, that minimizes the energy. Hubbard gave an algorithm for constructing this
ground state \cite{hubbard}; we will present two simple alternative constructions. The first gives the ground state for given $\nu=p/q$,  and
provides an intuitive understanding of why the energy is minimized; the second starts from $\nu=1/q$ and constructs the ground states at all other filling factors iteratively
from this state as repeated condensations of quasiparticles---in addition to the ground 
states this gives the fractionally charged quasiparticles and makes
the Haldane-Halperin hierarchy construction manifest in the TT-limit.

\begin{figure}[h!]
\begin{center}
\resizebox{!}{30mm}{\includegraphics{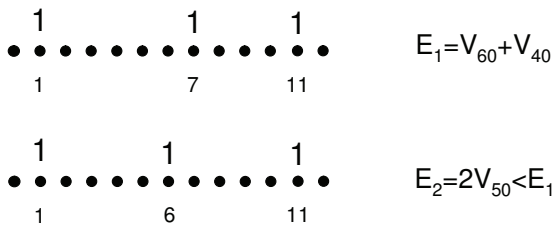}}
\end{center}\caption{\textit{{\small The interaction energy of one movable electron with two fixed 
electrons is minimized when the distances to the two fixed electrons differ by at most one lattice site.
}}}\label{concavity}
\end{figure}

The states obtained here are the ground states for any interaction that obeys the concavity condition
\begin{eqnarray}\label{vcond}
V_{k+1,0}+V_{k-1,0}> 2 V_{k0} \ \ \ ,
\end{eqnarray}
which implies, by iteration,
\be{gencond} V_{k+n,0}+V_{k-1,0}>
V_{k+n-1,0}+V_{k0}\ \ ,
\ee
for all $k,n>0 , \ k+n < N_s$. 
When, $L_1 \rightarrow 0$, $V_{k0}$ is simply the electrostatic interaction energy between 
two rings of circumference $L_1$ separated a distance $k2\pi/L_1$; hence
he concavity condition is fulfilled by a generic electron-electron
interaction when $L_1 \rightarrow 0$.
It implies that the interaction energy of one electron with two
other electrons that have fixed positions is minimized if the first
electron is as close to the midpoint between the fixed electrons as
possible, {\it ie} if the distances to the two fixed electrons
differ by at most one lattice constant, see Fig.
\ref{concavity}.
\footnote{Hubbard assumes, in addition, that $V_{k0}
\rightarrow 0 , \ {\rm as} \ k \rightarrow \infty $; this need not
be fulfilled when periodic boundary conditions are assumed as is the
case here.}

The crucial observation in obtaining the ground state is to realize that it is possible to minimize the energies of the
$k$:th nearest neighbors  separately for each $k$  for an interaction that obeys \pref{gencond}
\cite{hubbard}.

\begin{figure}[h!]
\begin{center}
\resizebox{!}{30mm}{\includegraphics{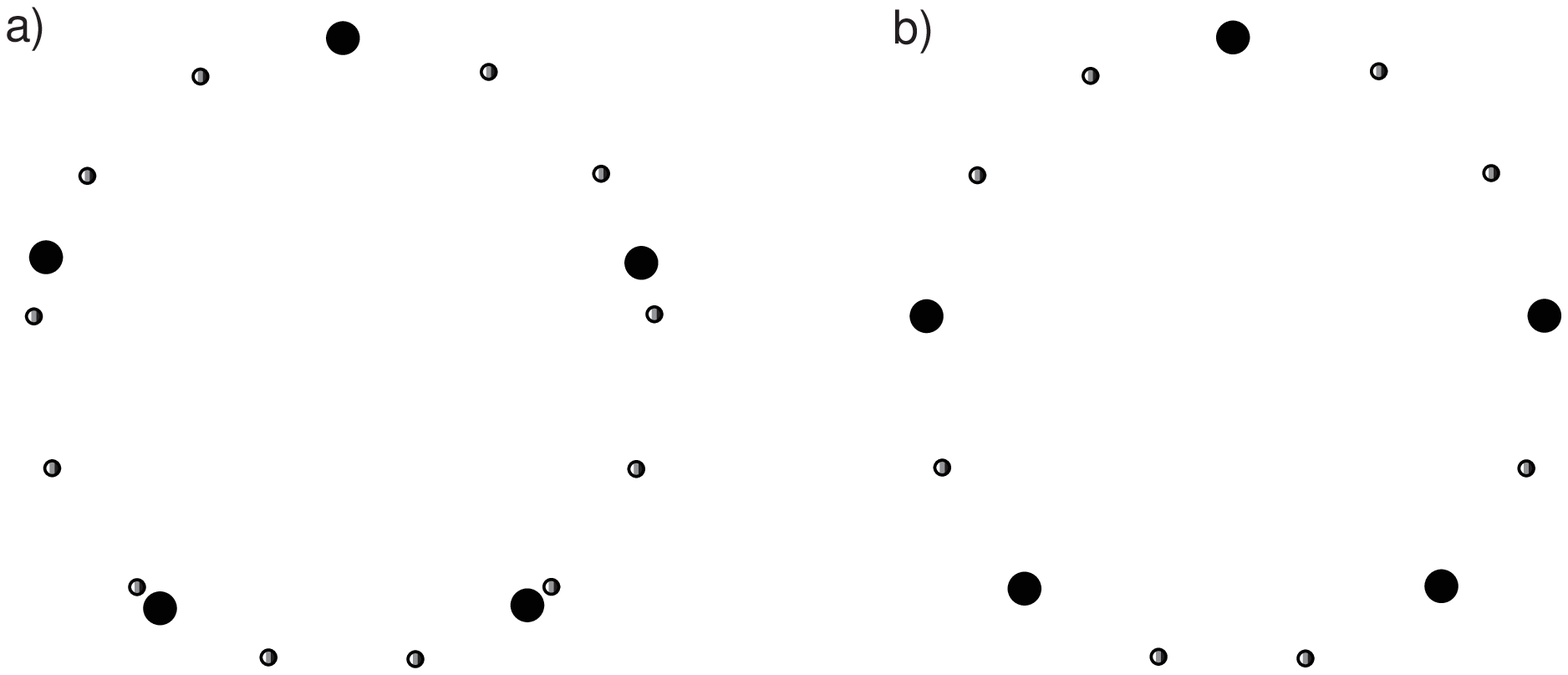}}
\end{center}\caption{\textit{{\small 
Ground state unit cell at $\nu=5/13$ by relaxation on circle. a) p=5 equidistant electrons on a circle with q=13 lattice sites;
b) each electrons moved to its closest site, unit cell is $(0_2101)_20_21$.}}}\label{circle}
\end{figure}

In  our first construction, the ground state is obtained by placing
the electrons equidistantly on a circle and then letting them relax
to the closest lattice sites. The ground state is periodic with a
unit cell of length $q$ containing $p$ electrons and this cell is
obtained as follows. Consider a circle with $p$ equidistant
electrons  and a lattice with  $q$ (equidistant) lattice sites as in
Fig. \ref{circle}.  Move each electron to its closest lattice site;
an electron that is equally far from two sites is moved to one of
these sites. \footnote{There is at most one such electron, since otherwise the period would be less than $q$.} 
The obtained configuration is the unit cell of the ground state, for
the proof we refer to Appendix \ref{appB}. An algebraic expression
for the unit cell is obtained by noting that electron $\alpha$ is at
site 
\be{uc1} i(\alpha) =I [\alpha/\nu] \ \ \ , 
\ee 
where $\alpha=
0,\dots, p-1$ and $I[x]$ denotes the integer closest to $x$. For example, for $\nu=5/13$, (\ref{uc1}) gives
$\{I[\alpha/\nu]\}=\{0,3,5,8,10,(13,\ldots)\}$ and hence the unit
cell is $1001010010100 \simeq (0_2101)_20_21$ in agreement with Fig.
\ref{circle}.  We use a chemical notation where the subscript
denotes the number of times the quantity is
repeated.\footnote{Hubbard uses a more compact notation where the
distance between consecutive ones is given, {\it eq},
$3^423^32^2=(0_21)_401(0_21)_3(01)_2$.} That the unit cell at
$\nu=p/q$ has length $q$ implies that the ground state is $q$-fold
degenerate. This is the  center of mass degeneracy of any state at
$\nu=p/q$ discussed in Sec. \ref{model} \cite{Haldane85PRL}. For the
quantum Hall states this is the topological degeneracy of the ground
state identified by Wen and Niu \cite{wen,wenreview}.

Incidentally, the initial configuration with equidistant electrons is clearly the ground state in the continuum problem when
there is no lattice---when there is a lattice the ground state is as similar to the continuum one as possible. Because of the periodicity, the picture of course
generalizes immediately to the full system consisting of $N$ unit cells and periodic boundary conditions: the ground state for $N_e$ electrons on $N_s$ sites
is obtained by starting with equidistant electrons and moving each electron to its closest site.

At $\nu=1/q$ the procedure gives the unit cell $0_{q-1}1$. Thus, the ground state is obtained by placing one electron on every $q$:th site; this obviously minimizes the
electrostatic repulsion. For odd $q$, these are the Laughlin fractions and the state with unit cell $0_{q-1}1$ was in fact proposed by Tao and Thouless in 1983
as an explanation of the fractional quantum Hall effect \cite{Tao83}. Although this state has a small overlap with the exact ground state (for large $L_1$),
and with the Laughlin state, it does in fact play an important role in the quantum Hall effect and we will call the crystal
ground states at general $\nu=p/q$ Tao-Thouless (TT) states. We claim that the TT-states are QH-states in the sense that they are adiabatically
connected to the bulk QH states \cite{bk2,hierarchy}.

Note that the TT-states have a gap to excitations---the
lattice sites are fixed in space and there are no phonons in these crystals; the TT-states should not be confused with Wigner crystals, which have gapless excitations due to the broken translational invariance.
This is a further reason to call the states considered here TT-states rather than crystal states.

We will now present an alternative, iterative, construction of the TT-states that brings out the connection to the hierarchy of fractional
quantum Hall states \cite{hierarchyHaldane,hierarchyHalperin} and determines the fractionally charged quasiparticles. Let ${\bf C}^{(n)}, n=0,1,2, \dots$ be the unit cell for a
TT-state at level $n$; this level is defined by the iteration process
and is identical to the one in the hierarchy construction. At the first levels we have
\begin{eqnarray}\label{firstcells}
{\bf C}^{(0)}&=&0 \ \ , \ \ \  \nu_0=0  \nonumber \\
{\bf C}^{(1)}&=&0_{t_1-1}1   \ \ , \ \ \  \nu_1=\frac 1 {t_1} \ \ ,
\end{eqnarray}
where $t_1=1,2,\dots$.
The unit cells for the TT-states at level $n\ge 2$ are obtained iteratively as
\be{iterative}
{\bf C}^{(n)}&=&{\bf C}^{(n-1)}_{t_n}{\bf C}^{(n-2)}, \ \ \  \nu_n=\frac {t_np_{n-1}+p_{n-2}}{t_nq_{n-1}+q_{n-2}} \ \  \nonumber \\
{\bf C}^{(n)}&=&{\bf C}^{(n-1)}_{t_n-1}\overline {{\bf C}^{(n-2)}},
\ \ \  \nu_n=\frac {t_np_{n-1}-p_{n-2}}{t_nq_{n-1}-q_{n-2}} \ \ ,
\ee where ${\bf C}^{(n-1)}_{t}$ indicates that ${\bf C}^{(n-1)}$ is
repeated $t$ times; $\overline {{\bf C}^{(n-2)}}$ is the
complement of ${\bf C}^{(n-2)}$ in the unit cell ${\bf C}^{(n-1)}$,
{\it ie} $ {\bf C}^{(n-2)}\overline {{\bf C}^{(n-2)}}={\bf
C}^{(n-1)}$ and $\nu_n = p_n/q_n$.
At the second level we obtain:
\be{level2}
{\bf C}^{(2)}&=&\{0_{t_1-1}1\}_{t_2}{0} \ \ , \ \ \ \nu_2 =\frac 1 {t_1+\frac 1 {t_2}} \nonumber \\
{\bf C}^{(2)}&=&\{0_{t_1-1}1\}_{t_2-1}0_{t_1-2}1 \ \  , \ \ \ \nu_2 =\frac 1 {t_1-\frac 1 {t_2}} \ \ .
\ee
We can now connect to the original hierarchy construction.
As will be further discussed below, ${\bf C}^{(0)}=0$ is the quasihole,
and $\overline {{\bf C}^{(0)}}=0_{t_1-2}1$ is the quasielectron, with charges $e^*=\pm e/t_1$,  in the
ground state ${\bf C}^{(1)}=0_{t_1-1}1$. Thus the unit cells ${\bf C}^{(2)}$ at level two consists of $t_2$ (or $t_2-1$) 
copies of the unit cell at level one, followed by a
quasihole (or quasielectron) in the level one ground state---the new ground state at level two is a condensate of
quasiparticles in the level one ground state.

At the next level, level three, we find the unit cells
\be{level3C}
{\bf C}^{(3)}&=&\{\{0_{t_1-1}1\}_{t_2}{0}\}_{t_3}0_{t_1-1}1 \nonumber \\
{\bf C}^{(3)}&=&\{\{0_{t_1-1}1\}_{t_2}{0}\}_{t_3-1}\{0_{t_1-1}1\}_{t_2-1}0    \nonumber   \\
{\bf C}^{(3)}&=&\{\{0_{t_1-1}1\}_{t_2-1}0_{t_1-2}1\}_{t_3}0_{t_1-1}1 \ \     \\
{\bf C}^{(3)}&=&\{\{0_{t_1-1}1\}_{t_2-1}0_{t_1-2}1\}_{t_3-1}\{0_{t_1-1}1\}_{t_2-2}0_{t_1-2}1 \nonumber
\ee
and the corresponding filling factors
\be{level3nu}
\nu_3 &=&\frac 1 {t_1+\frac 1 {t_2 \pm \frac 1 {t_3}}} \nonumber \\
\nu_3 &=&\frac 1 {t_1-\frac 1 {t_2 \pm \frac 1 {t_3}}} \ \ \ .
\ee
Generalizing this, we find at level $n$, the filling factors
\be{fractions} \nu_n=\frac 1 {t_1+\frac {\alpha_2} {t_2+\frac
{\alpha_3} {  \frac {{\cdot_\cdot} }    {t_{n-1}+\frac {\alpha_n} {t_n} } }
}} \ \ \ ,
\ee
where $t_i=1,2,\dots$
and the corresponding unit cells are obtained using
\pref{iterative}. In this construction, $\alpha_i=+1\, (-1)$ if
${\bf C}^{(i)}$ is constructed with ${\bf C}^{(i-2)}$ ($\overline
{{\bf C}^{(i-2)}}$) in \pref{iterative}. A state at level $n$ is
uniquely characterized by the parameters $\{t_1, \alpha_2t_2,
\alpha_3t_3,\dots, \alpha_nt_n\}$. Equation \pref{fractions} is the
continued fraction form of the filling factors as given by the
hierarchy scheme, except that now also even denominators are
obtained. Restricting to $t_1=1,3,5,\dots$ and $t_i=2,4,6,\dots$,
for $i \ge 2$, \pref{fractions} is identical to Haldane's formula
for the filling factors \cite{hierarchyHaldane}, which is known to
give each odd denominator fraction once.

This construction gives each rational filling factor $0\le \nu \le
1$  and by inspecting the
unit cells one finds that they are such that the distances between
$k$:th nearest neighbors differ by one lattice constant at most,
hence they minimize the energy and are identical to the ground
states obtained by the relaxation procedure above. To be
precise, we have not proven that the unit cells obtained by the two
methods are always identical, but we have checked this in many
examples and are convinced that this is the case.

The interpretation of \pref{level2} as a condensate of
quasiparticles generalizes to arbitrary level $n$ in \pref{iterative}:
\be{qps}
{\bf C}^{(n-2)} \ \  {\rm and} \ \ \overline {{\bf C}^{(n-2)}}
\ee
are the quasihole and the quasielectron (which is which depends on the state) in the ground
state with unit cell ${\bf C}^{(n-1)}$, and the ground state at
level $n$, ${\bf C}^{(n)}$, is a condensate of these quasiparticles
in accordance with the hierarchy construction. That the proposed
quasiparticles, ${\bf C}^{(n-2)}$ and $\overline {{\bf C}^{(n-2)}}$,
have the expected charges $e^*=\pm e/q$ at $\nu=p/q$ follows from
the Su-Schrieffer counting argument \cite{Schrieffer}, see Appendix
\ref{appC}.  Furthermore, they are domain walls between the
degenerate TT ground states. This was noted by Anderson
\cite{anderson} and stressed by Su
\cite{su1984,su2} based on exact diagonalization studies on the torus.
The quasiparticles discussed here are the ones with the elementary charge
$e^*=\pm e/q$ and the QH states are the simple abelian ones; as pointed out
by Wen \cite{wenreview} other quasiparticles  may also in principle
condense to form more complicated ground states. 

The iteration formula \pref{iterative} gives the one quasiparticle excitations in the state with
unit cell ${\bf C}^{(n)}$ and shows that these are  the lowest energy excitations at the corresponding filling factors; for example, inserting $01$ in
the ground state with unit cell ${\bf C}^{(2)}=00101$ gives the quasielectron and it is the lowest energy state
at $\nu=(2t_2+1)/(5t_2+2)$.

The energy of a quasielectron-quasihole pair at $\nu=p/q$ can be calculated in the limit $L_1\rightarrow 0$. According to the discussion
above, the quasiparticles in a  TT-state with unit cell ${\bf C}^{(n)}$  are ${\bf C}^{(n-1)}$ and  $\overline {{\bf C}^{(n-1)}}$. Thus, a minimally separated particle-hole pair is obtained by the replacement
\be{ppnn}
{\bf C}^{(n)}={\bf C}^{(n-1)} \overline {{\bf C}^{(n-1)}} \rightarrow \overline {{\bf C}^{(n-1)}} {\bf C}^{(n-1)} \ \  ,
\ee
in the ground state. Note that the replacement in \pref{ppnn}
amounts to a translation of the unit cell ${\bf C}^{(n-1)}$ with periodic boundary conditions on the cell itself, thus it
creates two domain walls (with the expected charge). It can be shown, see Appendix \ref{appD}, that $\overline {{\bf C}^{(n-1)}} {\bf C}^{(n-1)}$
differ from ${\bf C}^{(n)}$ only in that one electron has been moved one lattice constant.

A  separated particle-hole pair is obtained by translating $s$ consecutive cells as in \pref{ppnn}, {\it ie} with periodic boundary conditions on
each cell separately, or, equivalently, on all $s$ of them together.
This is equivalent to
inserting a string of $s-1$ unit cells ${\bf C}^{(n)}$  between the particle and hole
\be{pp}
{\bf C}^{(n)}_s \rightarrow  \overline {{\bf C}^{(n-1)}} {\bf C}^{(n)}
_{s-1}  {\bf C}^{(n-1)} ,\
\ee
and moves $s$ electrons (one in each unit cell) one lattice constant in the same direction.

The replacement \pref{ppnn} or \pref{pp}  implies an ordering of the particle-hole pair. The opposite ordering is obtained by instead making
the reverse replacement
\be{pp2}
{\bf C}^{(n)}_s  \rightarrow  {\bf C}^{(n-1)} {\bf C}^{(n)}_{s-1} \overline{  {\bf C}^{(n-1)}}  \ \ .
\ee
Again the replacement (\ref{pp}) or (\ref{pp2}) amounts to a translation of ${\bf C}^{(n)}_s $ assuming periodic boundary conditions. 

It is shown in Appendix \ref{appD} that the energy of the separated particle-hole pair in (\ref{pp},\ref{pp2}) is
\be{energypair}
E_{\rm ph}(s)=\sum_{k=1}^{s-1} k\Delta _{kq} + s \sum_{s\le k\le N_s/2q} \frac{\Delta _{kq}}{1\!+\!\delta_{k,\frac{N_s}{2q}}\!\!}  \ \  ,
\ee
where
\begin{eqnarray}\label{delta}
\Delta _{kq} \equiv V_{kq-1,0} -2V_{kq,0}+V_{kq+1,0} \ \ .
\end{eqnarray}
$ \Delta _{kq}$ is the change in energy for an electron with its two neighbors $k$ unit cells away when the initial electron is moved one lattice constant.
This energy, which is the second derivative of $V_{k0}$, is positive due to the concavity condition  \pref{vcond}, which thus ensures the stability of the
TT-ground state to particle-hole formation. Note that $E_{\rm ph}(s)$ is independent of the numerator $p$ of the filling factor.

It can be shown that the nearest neighbor pair excitation $s=1$, which has energy
\be{nnenergy}
E_{\rm ph}(1)=\sum_{1\le k\le N_s/2q}\frac{\Delta _{kq}}{1\!+\!\delta_{k,\frac{N_s}{2q}}\!\!} \ \ ,
\ee
according to \pref{energypair},
is the lowest energy excitation at fixed filling fraction, see Appendix \ref{appD}.

An important quantity is the energy of an infinitely separated
particle-hole pair; this is what is measured in activated transport
and is a measure of the stability of a quantum Hall state. When the
separation of the particles goes to infinity, $s\rightarrow \infty$,
we find
\begin{eqnarray}
E_{\rm gap}\equiv E_{\rm ph}(\infty)=\sum_{k=1}^{\infty} k\Delta _{kq}  \ .
\end{eqnarray}
Note that the gap depends only on the denominator $q$, {\it ie} on the fractional charge $\pm e /q$, and not on the numerator $p$ in $\nu=p/q$.

This is natural in the sense that the denominator determines the charge of  the quasiparticles, but the
result is non-trivial since the properties of the ground states depend on both $p$ and $q$.
Furthermore, $E_{\rm gap}$ is a monotonic function that approaches
zero from above as the denominator $q$ increases.\footnote{This
statement is true for most imaginable interactions. A sufficient
(but not necessary) condition is that $\Delta_k$ (or its derivative)
is monotonic in $k$.}
In this context it should be mentioned that Halperin within the original hierarchy construction of fractional QH states predicted a gap that
is predominantly determined by $q$ and decreases monotonously with increasing $q$ \cite{hierarchyHalperin}.

\begin{figure}[h!]
\begin{center}
\resizebox{!}{64mm}{\includegraphics{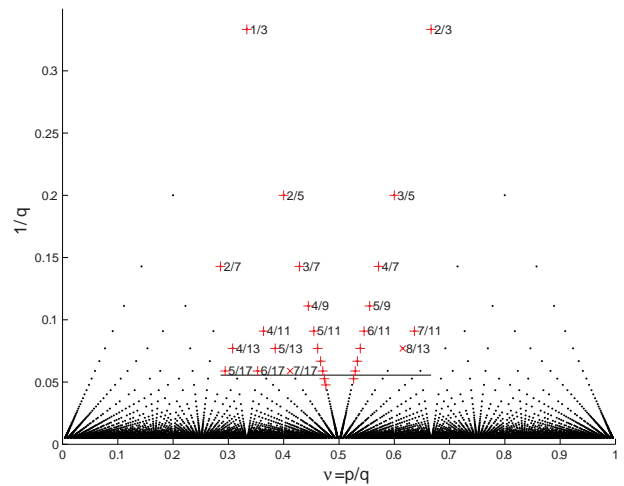}}
\end{center}\caption{\textit{{\small The Tao-Thouless states and their stability.
Each point $(p/q,1/q)$, $q$ odd, corresponds to a TT-state that we claim is adiabatically connected 
to a bulk QH-state. The gap, and hence the stability, increases with decreasing $1/q$. At the points 
marked by plusses or crosses indications of QH-states are observed \cite{pan}. The line is at 
constant gap and shows the range of the experiment in $\nu$. It is an approximate lower boundary  
in $1/q$ for the observed states.}}}\label{hierarchyfig}
\end{figure}

Thus we find that the gap to creating a separated quasiparticle pair decreases monotonously with increasing $q$.
This motivates  Fig. \ref{hierarchyfig} which shows $1/q$ for each  $\nu=p/q$,  $q$ odd \cite{hierarchy}. Each point corresponds
to a TT-state that we claim is adiabatically connected to  a bulk QH-state. The higher
up a point is, the larger is the gap in the corresponding TT/QH-state and the more stable is the state. The
points marked by plusses are filling fractions where a dip in the longitudinal resistance $R_{xx}$ is reported in the
experiment by Pan et al \cite{pan}. At the crosses, we infer a very small dip in the same data.
This experiment, which covers the range $2/7\le \nu \le 2/3$ marked in the figure,  is performed on the highest available
mobility samples and exhibits the largest number of QH states.
We note that to a
surprisingly good approximation a dip in $R_{xx}$ is observed at a filling factor $\nu=p/q$ if, and only if,  $q\le q_0 \approx 17$.
This is in agreement with the gap, $E_{\rm gap}$, being independent of $p$.

\begin{figure}[h!]
\begin{center}
\resizebox{!}{60mm}{\includegraphics{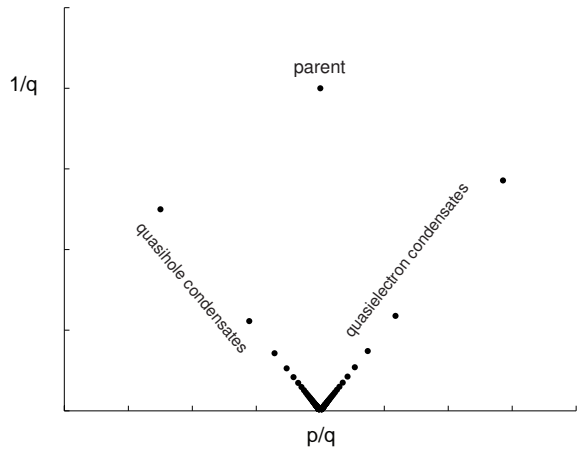}}
\end{center}\caption{\textit{{\small 
The hierarchy state at $\nu=p/q$ is the parent state for two sequences of daughter
states formed by condensation of quasielectrons and quasiholes in the parent state.These sequences approach $p/q$, with decreasing gap $\sim 1/q$,
from above and below respectively. Repeating this construction {\it ad infinitum} gives the fractal  structure in Fig. \ref{hierarchyfig}. }}}\label{relatives}
\end{figure}

The structure shown in Fig. \ref{hierarchyfig} is a fractal,
self-similar, one: enlarging any region of $\nu$ reproduces the
original figure \cite{mani,goerbig}. This fractal structure is  connected to the
hierarchy construction of fractional QH states. The TT-states and
their quasiparticles obtained when $L_1 \rightarrow 0$ makes the
hierarchy construction of fractional QH states manifest as discussed
above. According to \pref{iterative}, for each state, the parent
state, condensation of quasiparticles gives rise to two sequences of
daughter states with filling fractions approaching that of the
parent state from above and below respectively and with decreasing
$1/q$, see Fig. \ref{relatives}. This holds for each $\nu=p/q$ and
explains the fractal structure in Fig. \ref{hierarchyfig}. For a
more complete discussion of the connection to the hierarchy theory,
including the connection to the global phase diagram \cite{global,lutken}, we
refer to Ref. \onlinecite{hierarchy}.

Using Fig. \ref{hierarchyfig} we can predict what QH-states are next in line to be discovered when higher mobility samples become available. For example, in the region $2/7\le \nu \le 2/3$ we first expect, in addition to 7/17 and 8/13 included above 
and new Jain states  at $p/(2p\pm 1)$, states at
10/17, 11/17, 6/19, 7/19 and 8/19.

\subsection{Examples}\label{examples}

We here give explicit examples of TT ground states and quasiparticles for prominent filling fractions.

\subsubsection{Laughlin/Jain fractions}

For the Jain sequences $\nu=p/(2mp+1)$ that for fixed $m$ approach
$\nu=1/2m$ from below as $p$ increases, the unit cell of the
TT-state is $0_{2m}1(0_{2m-1}1)_{p-1}$; in the hierarchy notation
this corresponds to $\{t_1,\alpha_2 t_2,\dots, \alpha_p
t_p\}=\{2m+1,-2,-2\dots,-2\}$. Thus the states in the Jain sequences are those in the
hierarchy where all but the first condensate has maximal density. At $\nu=1/2m$, the
unit cell is $0_{2m-1}1$. Explicitly for the experimentally most
prominent sequence, $m=1$, the unit cells  are \be{TTjain}
1/3&:& \  001 \nonumber \\
2/5&:& \  00101 \nonumber \\
3/7&:& \  0010101 \nonumber \\
4/9&:& \  001010101 \nonumber \\
5/11&:& \  00101010101 \nonumber \\
\dot \nonumber \\
\dot \nonumber \\
\dot \nonumber \\
1/2&:& \ 01 \ \ \ .
\ee
By taking the particle-hole conjugate, $0\leftrightarrow 1$, of the states in \pref{TTjain} one obtains the Jain series $\nu=p/(2p-1)$
that approaches $1/2$ from above.

We now turn to the fractionally charged quasiparticles. Consider to start with the TT-state at $\nu=1/3$.  According to \pref{firstcells}, its unit cell is ${\bf C}^{(1)}=001$
and the quasiparticles are ${\bf C}^{(0)}=0$ and   $\overline{{\bf C}^{(0)}}=01$. A quasihole with charge $e/3$
is created by inserting ${\bf C}^{(0)}=0$  somewhere:
\be{qh1/3}
&&001001001001001001001001001001001001001 \\
&&00100100\underline{0}1001001001001001001001001001001 \nonumber \\
&&00100100\underline{0}100100100100\underline{0}100100100\underline{0}1001001\underline{001} \ . \nonumber
\ee
The first line in \pref{qh1/3} is the ground state; the second line has one extra 0 inserted---comparing to the ground state it is clear that this creates a domain wall
between two degenerate ground states. In the third line, three well-separated 0's are inserted, creating three domain walls. Comparing to the ground state in the
first line one sees that far away from the three 0's the state is unchanged and to maintain the size of the system the underlined unit cell $001$ has to be removed. In between the inserted 0's the state is a rigid translation of  the original ground state and thus indistinguishable from this state by a measurement that only refers to the translated state. Thus, the net change is that one electron has been removed and this charge is divided on the three well-separated domain walls which thus have charge $e/3$
each. This is of course nothing but the Su-Schrieffer counting argument \cite{Schrieffer}. The argument is also closely related to Laughlin's original argument for
the fractional charge, where a quasihole was created by adiabatic insertion of a flux quantum \cite{Laughlin83}---this corresponds to adding one empty site as
there is one site per flux quantum.

The quasielectron is created by inserting $\overline{{\bf C}^{(0)}}=01$:
001001001\underline{01}001001001001. Again this is a domain wall and the charge is, by the
counting argument $-e/3$. Note that particle-hole symmetry is manifest; this is not the case for the bulk wave functions.
It is straightforward to show that the excitations just given are the $L_1\rightarrow 0$ limits of the two-dimensional bulk
quasiparticles constructed by Laughlin,  see Appendix \ref{appE}.

It follows from Sec. \ref{solution} that inserting or removing 01 create quasiparticles in all the TT ground states in the
Jain sequence $\nu=p/(2p\pm 1)$ that approaches $1/2$
as $p$ increases. For example, a quasielectron at $\nu=2/5$ is obtained as:
\be{qh2/5}
&&00101001010010100101001010010100101\nonumber \\
&&0010100101\underline{01}0010100101001010010100101  \ \ . \ee For
Jain sequences with general $m$, starting from the Laughlin state
$\nu=1/(2m+ 1)$, quasiholes (quasielectrons) are created by
removing (inserting) $0_{2m-1}1$. In all cases, the Su-Schrieffer
counting argument gives the expected charge, $\pm e/(2mp +1)$, for
these quasiparticles and they are the $L_1 \rightarrow 0$ limits of the Laughlin and Jain
quasiparticles in the two-dimensional bulk system, see
Appendix \ref{appE}.

\subsubsection{Non-Jain fractions}\label{nonjain}

Until recently, all experimentally  well-established fractional quantum Hall states where for Laughlin-Jain filling fractions. However, in 2003 Pan {\it et al} reported a new set of states
in ultra-high mobility samples \cite{pan}. Preliminary indications of such a state at 4/11 was reported by Goldman and Shayegan \cite{goldman}.
Fractional quantum Hall states were seen at the following odd denominator filling factors
$\nu=4/11,7/11,4/13,5/13,(8/13),5/17, 6/17$ and $(7/17)$ (the ones in parenthesis  are inferred by us from the data in Ref. \onlinecite{pan} but were not claimed in this reference). \footnote{The
experiment shows similar features, a dip in the longitudinal resistance, at even denominator fractions, {\it eg} at $\nu=3/8,3/10$---these are presumably not QH states, see Sec. \ref{othereven}.}
Using the methods above, we readily find the TT ground states and the quasiparticles with charge $e^*=\pm e/q$ at these filling fractions.
In Table \ref{tablepan} the unit cells are given in the hierarchy form, {\it ie} as in \pref{iterative}. The anti-quasiparticle, which is not given in the table, is
obtained by taking the complement of the quasiparticle in the given ground state unit cell, according to \pref{qps}.
The unit cell for given $\nu=p/q$ is most easily obtained using the
relaxation method, \pref{uc1}; it  is then easily transformed to the hierarchy
form by identifying its parent using Figs. \ref{hierarchyfig} and \ref{relatives}. 
The parent is one of the two nearest neighbors whose denominator is smaller than $q$. {|tt Check this}
\begin{table}
\renewcommand{\arraystretch}{1.3}
\caption{Ground states  and quasiparticles in the TT-limit for the odd denominator fractions observed by Pan {\it et al} \cite{pan} that do not
belong to the Jain sequence. } \label{tablepan} \centering
\begin{tabular}{c|c|c}
    \hline
 $\nu$  &  ground state & quasiparticle    \\
    \hline
    \hline
4/11 & $ \{0_21\}_301$ & $0_21$ \\
7/11& $\{1_20\}_310$ & $1_20$ \\
4/13& $\{0_21\}_40$   &$0_21$  \\
5/13& $\{0_2101\}_20_21$ & $0_2101$ \\
8/13&$\{1_2010\}_21_20$ & $1_2010  $\\
5/17 &$\{\{0_21\}_20\}_20_21$ &    $ \{0_21\}_20 $   \\
6/17&$\{0_21\}_501$  & $0_21 $\\
7/17&$\{0_2101\}_301$& $0_2101 $\\
    \hline\end{tabular}
\end{table}

\subsection{Emergent Landau levels and composite fermions}\label{emergent}

We have seen that in the TT-limit  the TT ground
states give  a microscopic realization of the  original hierarchy
construction of fractional quantum Hall states. Today the
alternative construction of quantum Hall states in terms of
composite fermions has become the dominant framework and one may ask
how the TT-construction is related to this. We believe, as argued by
Read \cite{read90} and by Blok and Wen \cite{blokwen} but contrary to the opinion of
Jain \cite{jainrev1,jainrev2,jainbook}, that the original hierarchy construction and
composite fermions are just two different ways to view the same
phenomena. In our approach, this is supported by the fact that both
the hierarchy construction and the emergent Landau level structure
is manifest in the TT-limit and that the TT-ground states and the
quasiparticles are the $L_1 \rightarrow 0$ limits of Jain's
composite fermion wave functions.

\begin{figure}[h!]
\begin{center}
\resizebox{!}{30mm}{\includegraphics{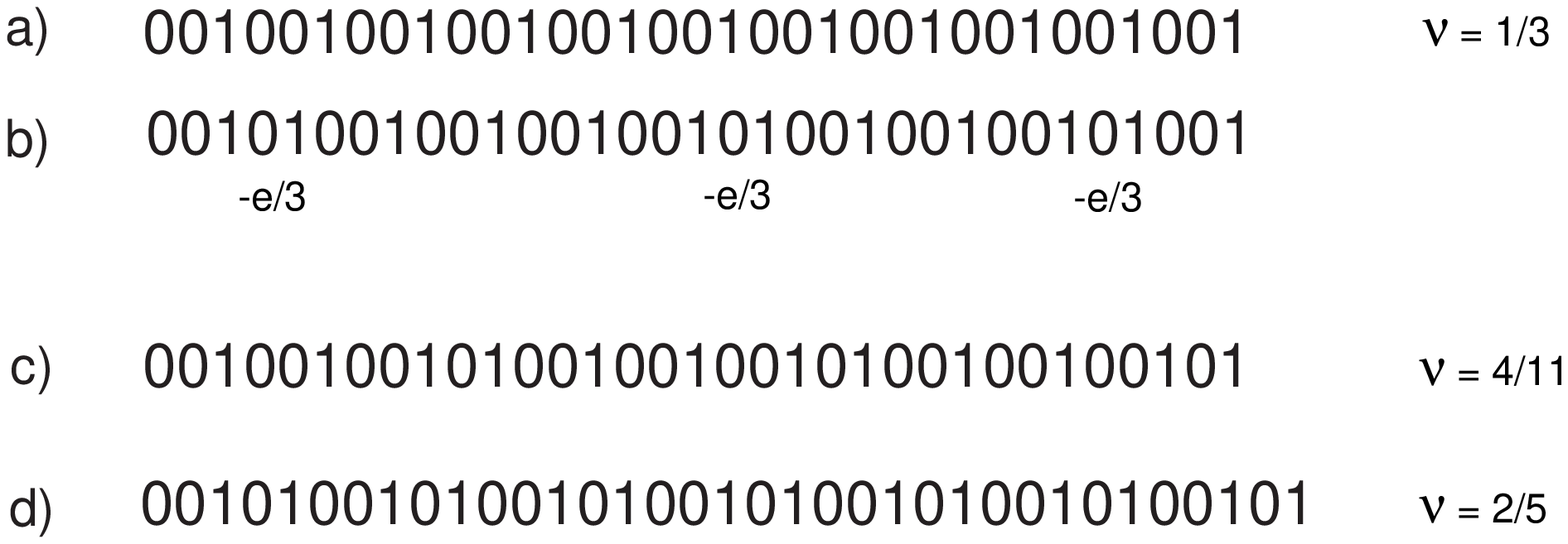}}
\end{center}\caption{\textit{{\small Quasiparticles and emergent Landau levels in the TT-limit. $(a)$ The
TT-ground state 001 at $\nu=1/3$; in $(b)$ three quasielectrons with charge $-e/3$ are created by
inserting 01 in three places. When one 01 is inserted for
every third unit cell 001, the TT-ground state at $4/11$ is obtained $(c)$. In $(d)$ the 2/5 ground state
is obtained by inserting one quasielectron for every unit cell 001. The ground states in $(c,d)$ can be
interpreted as filling emergent Landau levels.}}}\label{emergentLL}
\end{figure}

According to the composite fermion idea, fractional QH states are
created when composite fermions fill an integer number of Landau
levels in a reduced magnetic field. This implies the existence of an
emergent structure of effective Landau levels within the lowest
Landau level; such a structure is indeed seen in numerical
studies \cite{emergent,jainrev1}. We here describe how this comes about in the TT-limit.
To be specific we imagine that we start at
filling factor $\nu=1/3$ with the state  $\{0_21\}_{N}$, {\it
ie} the ground state has $N$ unit cells. When the magnetic
field $B$ is decreased, quasielectrons with charge $-e/3$ are
created by inserting 01. One such quasielectron can be inserted in
$N$ different equivalent places; these $N$ degenerate states
form an effective Landau level for these quasiparticles. When $B$ is
further decreased more quasielectrons are added to this effective
Landau level;  the quasielectrons repel eachother and are hence
pushed as far apart as possible. Eventually, one 01 has been added
per unit cell 001, the effective Landau level is filled and one has
reached the state  $\{0_2101\}_{N}$, {\it ie} the 2/5 state with
unit cell 00101, see Fig. \ref{emergentLL}.

Of course, when decreasing the magnetic field and adding quasielectrons 01 to the 1/3 TT-state many other, infinitely many in fact, TT-ground states are
obtained before the 2/5 state is reached, see Fig. \ref{hierarchyfig}.  Among them are the states with unit cells $\{0_21\}_{t_2}01$, see Fig. \ref{relatives};
these are just as 2/5 obtained by a condensation of quasielectrons in the 1/3 state albeit with a lower density of quasielectrons. However, these states
have a filling factor with larger denominator and hence a smaller gap and are thus less stable to disorder and may not form in a given sample.

Continuing decreasing the magnetic field  creates
quasielectrons in the $2/5$-state by adding 01 which have charge $-e/5$ (adding 01 just before reaching 2/5 removes quasiholes of charge $e/5$). After having added
01 an additional $N$ times one reaches the 3/7 ground state with unit cell 0010101. Continuing the process gives all the TT ground states in the Jain sequence
approaching $\nu=1/2$.\footnote{As well as states at all fractions in between just as was the case when going from 1/3 to 2/5.}
Note that whereas the operation of inserting 01 is one and the same throughout this process, its interpretation in terms of quasiparticles
varies: Near $p/(2p+1)$, adding 01 creates quasielectrons with charge $-e/(2p+1)$ and
when approaching  $(p+1)/(2p+3)$, adding 01 removes quasiholes of charge $e/(2p+3)$.

In the discussion above we started from the 1/3 state. However, due to the self-similar fractal structure established for the TT-ground states, it is clear that the
procedure is general---it applies  {\it mutatis mutandis} to any fraction $\nu=p/q$.

\section{Transition to bulk}\label{bulk}

We have seen in the previous section that the TT-limit $L_1\rightarrow 0$ is remarkably simple but still rich and non-trivial. However, this
is a limit that is most likely impossible to realize experimentally.  Its raison d'$\rm{\hat{e}}$tre lies in what it says about
the experimentally realizable two-dimensional bulk limit, $L_1 \rightarrow \infty$,---this is the topic of this section.

We propose that the TT-ground state is the $L_1\rightarrow 0$ limit
of a two-dimensional bulk QH state for any $\nu=p/q$, $q$ odd. If
this QH state is the ground state (in the bulk) then it is
adiabatically connected to the TT-state---there is in general no phase
transition from the TT-state as $L_1$ goes from 0 to
$\infty$.\footnote{Of course more complicated scenarios are possible
where there are intermediate phases and the QH phase is reentrant.} This statement may need
clarification. When $L_1$ changes, the size of the physical
system changes; normally when one talks about adiabatic continuity
one has in mind that a parameter in the hamiltonian changes.
However, all that is happening when $L_1$
changes is that the matrix elements $V_{km}$ change. Thus for any
$L_1$, we can think of the problem as the infinite
two-dimensional QH system with an interaction that depends on
$L_1$. This interaction is unusual: for finite $L_1$ it is
anisotropic, it becomes purely electrostatic in the TT-limit, 
and it cannot, presumably, be written as a real space interaction $V(r,L_1)$.
However, in the limit
$L_1\rightarrow \infty $ it approaches the chosen isotropic
electron-electron interaction, whereas it is exactly solvable in the
TT-limit. Thus, we have the standard
situation when adiabatic continuity can be discussed. In passing, we
note that this point of view shows that the TT-states give an exact
solution to the two-dimensional QH-problem albeit with a peculiar
interaction.

We will present arguments for our proposal about adiabatic
continuity below; the strength of these arguments varies with $\nu$.
For the Laughlin states adiabatic continuity holds for a
short-range pseudo-potential interaction, for the Jain states it can be
strongly argued and for more general hierarchy states a case for it
is emerging. In our opinion, the overall evidence for the
correctness of the claim is convincing.

Apart from fractional QH states, there may of course be other ground states in the bulk. An example of this are Wigner crystals that are
expected at low filling factors, in particular at $\nu=1/q$, for $q$ small enough. In these cases, there is a phase transition from the TT-state
as $L_1$ increases.

The even denominator filling factors are special. For them we propose that there is always a phase transition from the gapped TT-state
as $L_1$ increases. The suggestion is based on: 1)  a detailed analysis of the
$\nu=1/2$ case and 2) exact diagonalization of small systems. For $\nu=1/2$, we find a phase transition to a gapless Luttinger
liquid that can be identified with a version of the Rezayi-Read state \cite{rr}; when  $L_1$ increases further the state develops smoothly 
into the Rezayi-Read state that is believed to describe the system in the limit $L_1\rightarrow \infty$.  Exact diagonalization of all  $\nu=p/q$ with $q\le 11$ shows that for each filling
factor with even denominator there is an abrupt change in the ground state, where its quantum numbers change,  as $L_1$
increases, whereas there is no such change for any odd denominator filling factor \cite{bk2}.

Below we first discuss the odd denominator hierarchy states, concentrating on the Laughlin and Jain fractions; results for more general hierarchy
fractions are summarized. We then turn to even denominator fractions;  we discuss
the half-filled Landau level in detail and comment on general such fractions and on the difference between $\nu=1/2$ and $\nu=p/(2p+1)$.
The TT-limit is useful also to analyze non-abelian states; we end by discussing the Moore-Read pfaffian state at $5/2$ as an example of this.

\subsection{Hierarchy states}\label{hierarchybulk}

We begin by comparing the qualitative properties of the TT-states and the bulk hierarchy states. First of all, both
have a gap to excitations. Note that the crystalline TT-states have no phonons: once the boundary conditions, {\it ie} the flux through
the torus, are fixed the lattice sites are fixed in space. Thus the TT-states are not the Wigner crystals expected
in the quantum Hall system at low filling factor; these would have gapless phonons.
Second, the TT-state and the bulk hierarchy state at $\nu=p/q$ both have quasiparticle
excitations with charge $e^*=\pm e/q$. A further similarity is that  the ground state degeneracy on the torus in both
cases is $q$, but this is rather trivial as it is true for any state at $\nu=p/q$.

On the torus, the symmetries are the
magnetic translations and the TT and Laughlin-Jain states have the same quantum numbers, $K_{\alpha}$,  with respect
to these.  This follows since the former are the $L_1 \rightarrow 0$ limit of the latter, see Appendix \ref{appE}.

The TT-state is inhomogeneous,  whereas the
bulk hierarchy state should be homogeneous---a property that is normally considered fundamental for the quantum
Hall fluids. However, it should be noted that imposing periodic boundary conditions on the problem of electrons
moving in a perpendicular magnetic field, {\it ie} considering the problem on a cylinder or a torus, implies that the continuous
translational invariance present on the infinite plane,  is broken, in one direction, to a discrete invariance, translation by one lattice constant $2\pi/L_1$, along
the cylinder, see Appendix \ref{appA}. As a consequence, for any state on a finite cylinder or torus there will be
ripples in the density that disappear completely only as the size of the torus goes to infinity  \cite{haldanebook}.

Furthermore, it should be noted that if one takes the alternative "adiabatic" point of view where the
TT-limit is just a change in the interaction $V_{km}$ in the infinite two-dimensional system (where the lattice constant vanishes), then the TT-state is
homogeneous.

\subsubsection{Laughlin-Jain fractions}\label{laughlinjain}

For the Laughlin and Jain filling fractions, $\nu=p/(2mp+1)$,  explicit wave functions are known for the ground states
as well as  for the quasiparticle excitations---these allow us to further investigate the connection to the TT-states.

Rezayi and Haldane noted in 1994 that the $L_1\rightarrow 0$ limit of Laughlin's wave function (at $\nu=1/3$)
is the TT-state with unit cell 001 \cite{Haldane94}. This
analysis can be generalized to general Jain states and also to quasiparticle excitations thereof.  Taking the $L_1 \rightarrow 0$ limit of these
wave functions, one finds the TT-states and the quasiparticles of Sec. \ref{solution}  \cite{bk2}. One direct consequence of
this is that the TT-state has the same quantum numbers $K_\alpha$ as the Laughlin/Jain state.
The details of this analysis are given in Appendix \ref{appE}.

On the plane, the Laughlin state at $\nu=1/(2m+1)$ is the unique ground state to a certain short range interaction
and there is a gap to all excitations \cite{haldanebook,trugman,pokrovsky}.
Since this result on the plane is a consequence of the behavior of the wave function when two electrons approach each other
it is natural that it holds in other geometries as well.
This is indeed the case: For the short range interaction
on the torus the Laughlin state is the unique ground state (up to the center of mass degeneracy) and there is a gap to
all excitations for any circumference $L_1$. 
An explicit statement or proof of this has not, to the best of our knowledge, appeared, but it
is implicit in Ref. \onlinecite{Haldane94}.\footnote{However, there
is a confusing comment in Ref. \onlinecite{Haldane94} on observed gapless modes which seems to say that such bulk modes are
expected.} For a proof see Appendix \ref{appF}.

The transition from an inhomogeneous  charge density wave state to a homogeneous-looking state as $L_1$ grows was studied
in \cite{Haldane94} for the Laughlin $\nu=1/3$ state and a rapid crossover was found at $L_1 \sim 6$. This
crossover was recently studied numerically in detail by Seidel {\it et al} \cite{Lee05}. They confirm that there is no phase transition
but an exponentially fast crossover to a virtually homogeneous state. That the Laughlin state spontaneously breaks the
discrete translational symmetry on the thin cylinder has been rigorously proven by Jansen {\it et al} \cite{Jansen}.

To conclude, we consider it firmly established that the TT ground states are adiabatically connected to the bulk QH states for the
Laughlin fractions and that a strong case has been made for this being true also for the Jain fractions.

\subsubsection{Non-Jain hierarchy fractions}\label{nonjain2}

For the fractional QH states observed by Pan {\it et al} \cite{pan} that are not of  the Laughlin-Jain type discussed above
no agreed upon wave functions exist. However,
recently wave functions were constructed for all filling fractions in the hierarchy scheme that are obtained as repeated condensates of
quasielectrons (as opposed to quasiholes) \cite{hierarchy,conformal}. These wave functions are obtained by a natural generalization of the conformal field theory
construction of Jain's composite fermion wave functions \cite{hans}. The $L_1\rightarrow 0$ limit of the proposed wave functions are the TT ground
states and  the wave function at 4/11 has a large overlap with the exact ground state for small systems. We take this as a strong indication that
the non-Jain hierarchy states in the bulk are also adiabatically connected to the TT ground states. Details of this conformal field theory
construction of hierarchy wave functions will be presented elsewhere \cite{conformal}. Wave functions at these fractions can also be constructed
within the hierarchy scheme as condensates of quasiparticles \cite{greiter94}  and as fractional QH states of
composite fermions \cite{JainGoldman,Quinn,jain4over11}. For further discussion of these states, see Ref. \onlinecite{lopez2,wojs,scomp2,pashitskii}.

\subsection{Even denominator fractions}\label{even}

So far we have discussed odd denominator filling factors and argued that the TT-states that are the ground states in the
TT-limit are adiabatically connected to the two-dimensional bulk QH hierarchy states.  In this section we
discuss the even denominator filling factors. We begin by considering the half-filled Landau level; this case has been analyzed in
detail and we believe it is representative for other gapless even denominators. (The gapped state at $5/2$ is discussed
in Sec. \ref{pfaffian}.)

\subsubsection{Half-filled Landau level\label{halffilled}}

In the half-filled lowest Landau level,
$\nu=1/2$, a metallic, {\it ie}, gapless state is observed \cite{jiang}. The composite fermion explanation for this behaviour is that
each electron binds two flux quanta, thereby removing all the external magnetic flux. The state is then supposedly described
by non-interacting, or weakly interacting,  composite fermions in zero magnetic field, {\it ie}, by a free two-dimensional Fermi gas.
This is also the picture
that emerges from the mean field theory where the statistical magnetic field completely cancels the external magnetic field \cite{hlr}.
These descriptions  successfully explain experimental data and constitute the main evidence for the
correctness of the composite fermion idea. The microscopic wave function of composite fermion type describing the $\nu=1/2$ state
was provided by Rezayi and Read \cite{rr}---on the torus it reads \cite{rrhaldane}
\be{rr}
\Psi_{RR}  = {\rm det_{ij}}[e^{i{\bf k}_i\cdot {\bf R}_j}] \Psi_{\frac 1 2}  \ \ \ ,
\ee
where ${\bf R}_j$ are the guiding center coordinates
and $\Psi_{\frac 1 2}$ is the bosonic Laughlin state at $\nu=1/2$, see Appendix \ref{appA}. The wave function $\Psi_{RR} $
is not unique but depends on a set of parameters, "momenta",  $\{{\bf  k}_i \}$.  In the two-dimensional
bulk limit these are assumed to form a circular Fermi sea.

The TT-state at $\nu=1/2$ has unit cell 01, {\it ie} it is a state with one electron on every other site. This clearly minimizes the
electrostatic repulsion, and hence is the ground state, in the TT-limit $L_1 \rightarrow 0$.
As any TT-state it has a gap to excitations
and it has quasiparticles
with charge $\pm e/2$. Clearly, this is very different from the gapless state that is observed in the two-dimensional bulk limit
and is believed to be described by \pref{rr}.

A gapless state does in fact exist also for small but finite $L_1$ and there is strong numerical evidence that this state
develops without a phase transition into the gapless two-dimensional state described by \pref{rr}. The first evidence for this
was obtained in a numerical study of the QH system on a thin cylinder \cite{bkdmrg}, using the density matrix renormalization group method \cite{white}.
At $\nu=1/2$, a sharp transition from the TT-state with a finite amplitude of the density oscillations to a virtually homogeneous
state was observed at $L_1 \sim 5$ (for a short-range interaction $V(\mathbf r) = \nabla^2 \delta (\mathbf r)$) and there were indications of gapless
excitations. This behaviour was dramatically different from other fractions such as $\nu=1/3$ where no transition was observed
as $L_1$ was varied---the amplitude of the density wave decreased continuously as $L_1$ increased.  The transition at $\nu=1/2$ takes
place in a region where only the first few shortest-range terms $V_{km}$ are important.   This made it possible to find an exact solution---ground state and low energy excitations---for a hamiltonian that is a very good approximation to the short-range interaction for $L_1 \sim 5$ \cite{bk1}.

We now describe the exact solution, for details see Appendix \ref{appG}.
Consider to start with a hamiltonian \pref{ham} that consists only of the shortest range hopping term, $V_{21}: \, 1001\leftrightarrow 0110$.
Define a restricted Hilbert space $\cal H'$ by requiring that each pair of nearby sites $2p, 2p+1$ contains exactly one electron, {\it ie} include the states
$\uparrow =10$ and $\downarrow=01$, but not 00 or 11.\footnote{If one instead groups sites $2p-1,2p$, one gets a restricted Hilbert
space  $\cal H'_T$ that is related to $\cal H'$ by a $T_2$ symmetry operation. The hamiltonian in \pref{ham} have vanishing matrix elements
between the states in $\cal H'$ and $\cal H'_T$ since $T_2$ is conserved and  these spaces have no common states except for the
two TT-states with unit cell 01.} Note that the interaction $V_{21}$ preserves the restricted Hilbert space $\cal H'$.
In spin notation, $1001\leftrightarrow 0110$ becomes $\uparrow \downarrow \leftrightarrow \downarrow \uparrow$
and hence the model with the shortest range hopping term $V_{21}$ is, within $\cal H'$, equivalent to a nearest neighbour
spin-1/2 $xy$-chain,
\be{xy}
H_{xy}= V_{21}\sum_{p=0}^{N_s/2-1} (s^+_{p+1}s^-_{p}+s^-_{p+1}s^+_{p}) \ \ \ ,
\ee
where $s^+_p=c^\dagger_{2p}c_{2p+1}$, $|\uparrow \rangle=s^+_p|\downarrow \rangle$, and $V_{21}>0$. Note that $s^+$ creates a neutral particle, a dipole.
The hamiltonian \pref{xy} is diagonalized by mapping the spins to fermions, $\tilde d_k$,  via a Jordan-Wigner transformation---the hamiltonian then
becomes that of free one-dimensional fermions with nearest neighbor hopping.

Thus, within $\cal H'$, the shortest range hopping term $V_{21}$ gives a one-dimensional free Fermi gas of neutral particles. But
to make this a reasonable approximation to \pref{ham} for small but finite $L_1$ we must consider also the short-range electrostatic terms. Including
the two shortest range such terms, $V_{10}$  and  $V_{20}$, with relative strength $V_{10}=2V_{20}=2\alpha $ one finds that all states in
$\cal H'$ have the lowest possible electrostatic energy. There are states not in $\cal H'$ (or $\cal H'_T$) with the same electrostatic energy but a strong
case can be made that the  hopping term $V_{21}$ creates a gap to these ($\cal H'$ contains the most hoppable states) \cite{bk1}; this is also supported by the 
numerics discussed below. Thus,  the low energy
sector for the short-range hamiltonian
\be{shortham}
H_{ sr}&=&\alpha \sum_{p=0}^{N_s-1} \hat n_p (2\hat n _{p+1}   +  \hat n _{p+2} ) \nonumber \\
&+&V_{21}\sum_{p=0}^{N_s/2-1} (s^+_{p+1}s^-_{p}+s^-_{p+1}s^+_{p}) \ \ ,
\ee
where $\hat n_p=c^\dagger _p c_p$, is contained in $\cal H '$ and is that of a free one-dimensional Fermi gas of neutral particles.
The ground state is a filled Fermi sea
\be{groundstate}
 \ket{GS}=\prod_{|k|>\frac \pi 2}
  \tilde{d}^\dagger_{k}\ket{010101\ldots} \ \ ,
\ee
and excitations out of this sea gives gapless neutral excitations. This provides an exact and explicit mapping of the low-energy sector of a system
of interacting electrons in a magnetic field onto a system of non-interacting neutral particles, {\it ie} onto free particles that do not interact
with the magnetic field.

We note that dipoles have many of the properties expected of composite fermions, most notably that they do not couple to the magnetic field. 
Earlier approaches to a microscopic theory in terms of dipoles were made by Read \cite{readearly, read.5}, Murthy  and Shankar \cite{ms}, Pasquier 
and Haldane \cite{pasq}, Lee \cite{dhlee} and by Stern {\it et al} \cite{stern}.

An important feature of the exact solution is that the number of dipoles is not conserved. These particles are neutral and their number is not tied to 
the number of electrons in the system as is the  case for the composite fermions. For example, there are excitations where a single dipole is added to, 
or removed from, the Fermi sea. Moreover, the number of dipoles making up the Fermi sea  in \pref{groundstate}  is determined dynamically, {\it ie} 
by filling the negative energy dipole states; this leads to there being one
dipole for every second electron. As this is a consequence of energetics,  the number of dipoles  presumably changes as $L_1$ increases.
Assuming the ground state develops into a two-dimensional Fermi gas of dipoles as $L_1\rightarrow \infty$, this suggests that the number of dipoles, and hence the 
Fermi momentum, is determined dynamically rather than being simply given by the number of electrons. In this context we note that 
the Fermi momentum measured in surface acoustic wave experiments \cite{Willettreview} disagrees with the mean field theory result which follows from that the number of 
composite fermions in the Fermi sea is equal to the number of electrons \cite{hlr}.

The solvable case is a good approximation for a small finite $L_1$. For the real-space short-range interaction  $V(\mathbf r) = \nabla^2 \delta (\mathbf r)$,
$V_{10}=2V_{20}=2\alpha$ corresponds to $L_1=2\pi/\sqrt{2\ln 2}=5.3$. The short-range hopping term is then $V_{21}=\frac 3 8 \alpha$,
whereas  the leading ignored terms are small: $V_{30}=\frac 9 {128} \alpha$ and $V_{31}=\frac 1 {32} \alpha$.  This is close to the solvable point.

Longer  range hopping terms in \pref{ham} will in general not preserve the subspace $\cal H'$. However, since there is a
gap to states not in $\cal H'$, the effect of small such terms can be included perturbatively in an effective spin-1/2 hamiltonian that acts within $\cal H'$.
To zeroth order the effective hamiltonian is
\begin{eqnarray}
H_{eff}=\sum_{p,n}
[V_{2n,1}(s^+_{p+n}s^-_{p}+s^-_{p+n}s^+_{p})
-\Delta_{2n}s^z_ps^z_{p+n}]\ ,\label{effham}
\end{eqnarray}
where $s^z_p=\frac 1 2 (n_{2p-1}-n_{2p})$ and $\Delta_{2n}$ is given in \pref{delta}\footnote{Note that $\Delta_{2n}$ need not be positive 
for finite $L_1$}. 
In particular, this effective hamiltonian, $H_{eff}$, contains an Ising term $s^z_ps^z_{p+n}$ that
changes the non-interacting Fermi gas to an interacting Luttinger liquid with interaction parameter $K \neq 1$. If such terms come
with large coefficients, they may cause a phase transition that opens up a gap; the results below strongly indicate that this
does not happen in this system. In higher order perturbation
theory, one gets increasingly more complicated spin-1/2 models containing renormalized quadratic as well as higher
order terms. However, our numerical calculations suggest that these higher order terms may be very small
for a range of $L_1$---we find that the
projection of the exact ground state of the Coulomb interaction on $\cal H'$ is virtually unity in the "solvable" region
$5.3 < L_1< 8$ as discussed below.

We have numerically identified the state obtained in exact diagonalization in the solvable region with the
exact solution \pref{groundstate}. We continuously
varied $V_{km}$ from the values that give the exact solution to values that correspond to a $L_1$ in this region and verified that the
ground state develops continuously without any level crossing. Whereas $H_{xy}$ gives a qualitative understanding,
we suggest that $H_{eff}$ gives a quantitatively very accurate description of the low energy sector within this region. 
Thus we conclude
that the exact solution is stable in a finite neighborhood of the solvable point where it develops into an interacting gapless
one-dimensional model, a Luttinger liquid.

However, the perturbative argument for this is of course restricted to hamiltonians close to the solvable one---when $L_1$
increases, hopping terms of increasing range and with large coefficients will appear---leading to a hamiltonian
that is very different from \pref{shortham} and that definitely does not preserve the restricted hilbert space $\cal H' $.
To investigate this region we compared the ground states obtained in exact diagonalization with the Rezayi-Read states \pref{rr} \cite{bk2}.

\begin{figure}[h!]
\begin{center}
\resizebox{!}{50mm}{\includegraphics{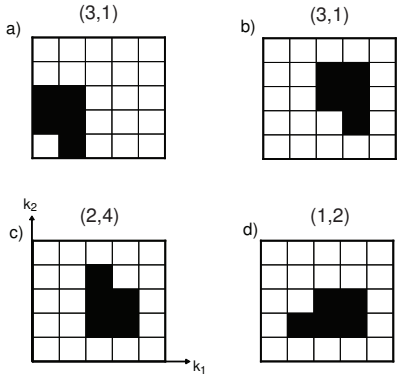}}
\end{center}\caption{\textit{{\small Examples of Fermi seas of momenta $\{k_{i\alpha}\}$ for $N_e=5$. For each sea the conserved 
momenta $(K_1,K_2)$ are displayed. $a)$ and $b)$  are related by a translation that corresponds to K-invariance, 
and have the same quantum numbers;  they describe the same state. 
$c)$ is one of three new states obtained from $a)$ by reflections in the $x$- and $y$-axes. These four states have different quantum numbers 
and are  degenerate ground states for $5.3<L_1<L_2$, {\it cf} Fig. \ref{phasediagrams}; translating the corresponding Fermi seas one step in the $x$-direction
gives the remaining four ground states. $d)$ is obtained from $c)$ by a rotation by $\pi/2$; it 
is one of the ground states when $5.3<L_2<L_1$.
}}}\label{boxes}
\end{figure}

The plane waves in \pref{rr} are  periodic on the torus, with our normalization this means that $k_{\alpha}$ are integers. 
A straightforward calculation shows that these momenta determine the conserved quantum numbers $K_{\alpha}$ of the state:
\be{}
K_{\alpha}=\sum_{i=0}^{N_e-1}k_{i\alpha} \ \ {\rm mod} ((1+\delta _{1\alpha})N_e) \ \ ,
\ee
where $k_{i\alpha}$ is the momentum of particle $i$. We 
represent a state \pref{rr} by displaying its set of momenta as in  Fig. \ref{boxes}. 
The Rezayi-Read state is invariant under a rigid translation of 
the momenta, $k_{i\alpha} \rightarrow k_{i\alpha} +(1+\delta _{1\alpha})n_\alpha$, where $n_\alpha$ are integers \cite{kinvariance} (this so-called 
$K$-symmetry was first noted by Haldane). Since the conserved 
quantum numbers are defined modulo $(1+\delta _{1\alpha})N_e$, see Appendix \ref{appA}, they are invariant under this translation.
The rigid translation $k_{i\alpha}\rightarrow  k_{i\alpha}+\delta_{1\alpha}$ changes the quantum numbers but corresponds to a translation of the center of mass only.

\begin{figure}[h!]
\begin{center}
\resizebox{!}{130mm}{\includegraphics{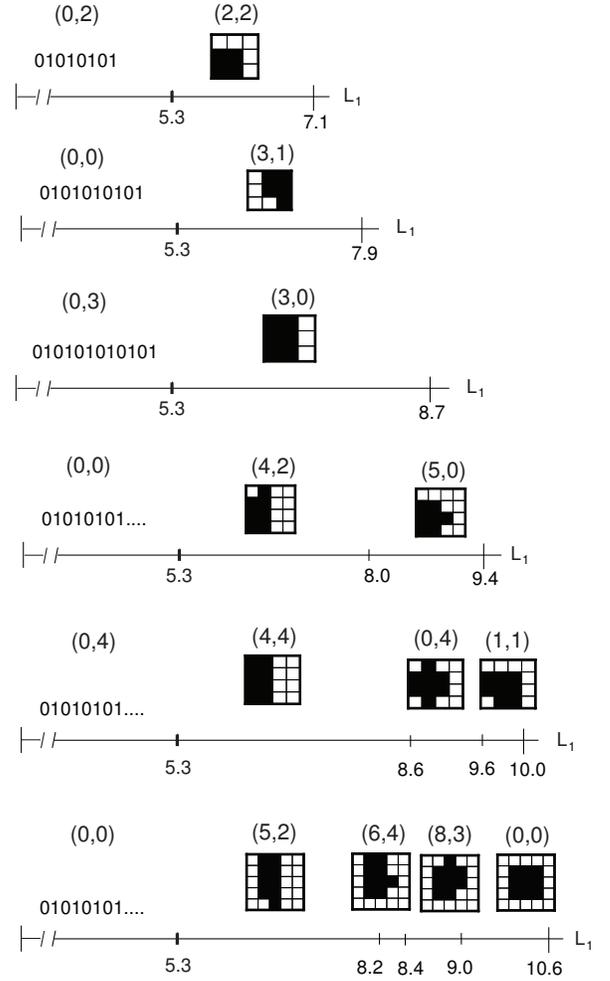}}
\end{center}\caption{\textit{{\small  Ground states and corresponding quantum numbers
$(K_1,K_2)$ at $\nu=1/2$ as a function of $L_1$ for  4 to  9
      electrons (from top to bottom) ($K_1$ is given ${\rm mod} (N_e)$). 
      The results are obtained in exact
      diagonalization, using an unscreened Coulomb
      interaction. Marked $L_1$ denote transitions to new ground states.  For $L_1 < 5.3$ the (virtually exact) ground state is  the TT-state, whose unit cell is 01
      when $L_1 \rightarrow 0$.  In each region $L_1 >5.3$, the state is identified with a  Rezayi-Read state with the
      displayed Fermi sea of momenta. The overlap with the Rezayi-Read state is around or above 0.99 for all $N_e$ (in all regions).  The phase diagram is symmetric about
      $L_1=L_2=\sqrt{4\pi N_e}$, which are the largest marked $L_1$ in the figure.}}}\label{phasediagrams}
\end{figure}

Fig. \ref{phasediagrams} shows results for the ground state at $\nu =1/2$ for various $L_1$ for $N_e=4-9$ electrons which interact
with an unscreened Coulomb interaction. For all $N_e$ there is a sharp transition from the TT-state to a new ground state at $L_1 \approx 5.3$. (In cases
where the ground state is degenerate, {\it cf} Fig. \ref{boxes}, only one representative is considered in Fig. \ref{phasediagrams}.)
As $L_1$ increases further there are additional transitions. The next transition occurs also at approximately the same $L_1$ for all  $N_e$;
this is particularly clear when an even/odd effect is ignored and
one considers only say the odd $N_e$.
There are additional transitions for larger $L_1$---but the number and positions of these depend on the number of particles. Each of the states at $L_1>5.3$ has a very large overlap
with a Rezayi-Read wave function for some choice of Fermi sea of parameters $\{{\bf k}_i \}$. Overlaps are around or above 0.99 and they are
essentially constant in each region. The Fermi seas develop in a systematic way from an elongated shape at small $L_1$
to a circular  one as $L_1$ increases. It is as circular as it can be when $L_1=L_2=\sqrt{4\pi N_e}$ for the small number of particles considered in Fig. \ref{phasediagrams}.

From the numerical results for a small number of electrons we extract the following interpretation
for an infinite number of electrons ({\it ie} for  $L_2 \rightarrow \infty$)  and varying $L_1$.  At $L_1 \approx 5.3$ there
is a phase transition from the gapped
TT-state, with unit cell 01 as $L_1 \rightarrow 0$, to a gapless state well described by the exact solution or equivalently by the
Rezayi-Read state with a special choice of momenta $\{{\bf k}_i \}$ forming an elongated Fermi sea. Perturbative corrections will turn this state into
a Luttinger liquid with $K\neq 1$ in the  region $5.3 <L_1<8$.
When $L_1$ increases further the Fermi sea deforms continuously  approaching a circular sea as $L_1 \rightarrow \infty$.
The transitions for $L_1 > 5.3$ in Fig. \ref{phasediagrams} are not phase transitions but rather level crossings to similar
states---such crossings are expected to occur in a gapless system. These transitions correspond to small changes of the momenta in the
Rezayi-Read wave function, typically only one momentum is changed at each transition. Consider a finite $L_1>5.3$ and $L_2 \rightarrow \infty$. This is a one-dimensional system, and
since it is obtained from the Rezayi-Read state that we have identified with the gapless exact solution by a
continuous change of momenta $\{{\bf k}_i \}$, we strongly believe that it is gapless. On general grounds, it should then be a Luttinger liquid---or possibly several Luttinger liquids.
Further support for it being gapless is obtained from
the fact that, as $L_1\rightarrow \infty$, it approaches the two-dimensional case which is generally believed to be gapless.
Of course, this is expected to be a gapless two-dimensional system---a free two-dimensional Fermi gas. It is an interesting
and unresolved question how the smooth transition from the Luttinger liquid at $L_1 > 5.3$ to the
two-dimensional gapless system occurs. In any case, our interpretation of the exact diagonalization results is that this transition
is smooth---there is no phase transition as $L_1$ varies above 5.3.

\subsubsection{Other even denominator fractions}\label{othereven}

For even denominator fractions other than $\nu=1/2$ we have much less to say; however, we suggest that the half-filled Landau level is typical and that a similar scenario
holds for other even denominator fractions.\footnote{Of course, we do not exclude other types of states at even denominators,
such as the Moore-Read state  discussed below.} There is some numerical evidence for this. We have performed exact diagonalisation, for various $L_1$, for all filling
factors  $\nu=p/q\le 1$ with $q\le 11$. All even  denominator fractions are similar to $\nu=1/2$ in that there are transitions from
the TT-state to new states with different quantum numbers as $L_1$ grows. For the odd denominator fractions, on the other hand, no such
transition is ever seen---the ground states develop smoothly from the TT-states. 

\subsubsection{Even versus odd denominators}\label{evenvsodd}

How can we understand, within the approach presented here,  that even and odd denominators are so different? To this question we have
no complete answer but we will offer some insights obtained from the case when the shortest range hopping term, $V_{21}$,  is included, {\it ie} when $L_1$ is small but finite.

Let us consider $\nu=1/2$ and the Jain sequence $\nu=p/(2p+1)$ that approaches this fraction as $p$ grows---why is 1/2 gapless whereas the others have a gap?
In the former case there is a phase transition from the gapped TT-state to a gapless state whereas in the latter case the gapped TT-state develops continuously
into the bulk quantum Hall system as $L_1 \rightarrow \infty$. To some extent this difference can be understood by comparing the TT-states.
As we have noted,
they are the states that minimize the electrostatic repulsion $V_{k0}$ and thus are the ground states as $L_1 \rightarrow 0$. When $L_1$
increases, hopping terms $V_{km}$, $m\neq0$, become important. Consider the first hopping term $V_{21}$ that enters as $L_1$ increases.
The TT-state at $\nu=1/2$ with unit cell 01 is annihilated by $V_{21}$, {\it ie} it is a non-hoppable state. There is a transition to a ground
state that is more hoppable, thus lowering the kinetic energy. The ground state in the exact solution \pref{groundstate} contains
the maximally hoppable basis state with unit cell 1001. For $\nu=1/2$, it is the competition between the electrostatic terms, that favor the TT-state, and the
hopping terms that favor other states, such as 1001, that leads to the phase transition at $L_1 =5.3$. For $\nu=1/3$, the TT-state has unit cell 001; by
construction this minimizes the electrostatic repulsion. But it is also the most hoppable state with respect to the hopping term $V_{21}$.
Thus, in this case there is no competition between electrostatic repulsion and hopping---they collaborate rather than compete---and hence
there is no phase transition. Of course, this does not explain why longer range hopping terms that are important for larger $L_1$ do not
cause a transition.  This argument extends to any filling fraction $\nu =p/(2p+1)$ in the Jain sequence
that approaches $1/2$. In spite of the fact that the TT-state for large $p$ looks very similar to the TT-state at $\nu=1/2$, which is non-hoppable, the
TT-state at   $\nu =p/(2p+1)$ is very hoppable. In fact, inserting a single extra hole in the $\nu=1/2$ TT-state gives a very hoppable state.

Thus we see that for $\nu=1/2$ there is a doubling of the "unit cell", $01 \rightarrow 1001$, that does not happen at 1/3.   This generalizes to other filling fractions
and we suggest that the most hoppable state at odd denominators is simply the TT-state with unit cell  ${\bf C}$, whereas at even denominators
it is the state with doubled unit cell ${\bf C}^T{\bf C}$, where ${\bf C}^T$ is the transpose of ${\bf C}$. This is in agreement with the doubling of the period in
the "parent state" for even denominators noted by Su \cite{su2}. It is an open question how these considerations relate to the explanation by
Tao and Wu \cite{taowu}.

We end this section by pointing out an intriguing possible relation to the Haldane conjecture for the presence of gaps in spin chains \cite{haldanegap}.
The gapless half-filled Landau level is, in the exact solution above, mapped onto a spin 1/2 system. The low-energy sector is obtained by grouping
the sites in pairs with one electron in each pair. Attempting a similar mapping of the low energy sector
at 1/3, when the leading hopping term $V_{21}$ is included, suggests grouping the sites in sets of three with one electron in each group. This gives
three states per group, and a mapping to a spin one chain. This leads us to speculate that the existence (absence) of a gap at odd (even)
denominators in the QH system is related to the Haldane conjecture for spin chains that says that integer spin chains have a gap whereas half-integer
chains are gapless.

The arguments given here are at present restricted to the QH system with a small, but finite, $L_1$ since only the shortest range hopping term $V_{21}$ is
included. However, in this context it is interesting to note that the ground state obtained at $\nu=1/3$ with a hamiltonian that only includes $V_{21}$,
without any electrostatic terms, has a 98\% overlap with the Laughlin state for six electrons at $L_1=7$ \cite{juha}. Of course, when the number of
particles, and $L_1$, increases this overlap will drop as longer range hopping terms become important.

\subsection{Non-abelian states---Moore-Read state at 5/2}\label{pfaffian}

So far we have seen that the gapped hierarchy states as well as the gapless $\nu=1/2$ state exist for small $L_1$. This limit is relevant also for
non-abelian states, where a simple understanding of the nontrivial degeneracies and fractional charges appear
\cite{we06,haldaneAPS,seidel06}.
As an example of this we discuss the Pfaffian state proposed by Moore and Read \cite{mr}. This gapped state is a
competitor to the gapless Rezayi-Read state \pref{rr} for a half-filled Landau level. It was suggested 
\cite{greiter} as a candidate for the observed
gapped $\nu=5/2$ state \cite{52exp}.\footnote{Ignoring
the two filled Landau levels,
this becomes effectively a $\nu=1/2$ problem albeit with a modified electron-electron interaction $V_{km}$ since only the
one-particle wave functions change.}   There is now numerical evidence that this suggestion is indeed
correct \cite{morf,rrhaldane,moller}.

The Moore-Read state is peculiar: It has a six-fold degeneracy on the torus rather than
only the two-fold implied by the filling factor, it has excitations with charge $\pm e/4$ rather than $\pm e/2$
as one would expect for a gapped state at half-filling and most surprisingly a state
with $2n$ quasiholes has degeneracy $2^{n-1}$ for fixed quasiparticle positions. Moreover, these quasiparticles have non-abelian fractional statistics.  These properties
were obtained by an intricate relationship to conformal field theory.  It has been proposed that the topological and non-abelian properties of these states
can be used for constructing a topologically-protected decoherence-free quantum computer \cite{Freedman}.
The Moore-Read state is the exact ground state of a certain local three-body interaction \cite{greiter,rrhaldane}. This is true on a torus for any $L_1$ and
can be used to show that the ground states in the TT-limit are the TT-states with unit cells $01$ and $0110$, {\it ie} the unique states where the distance
between all pairs of next nearest neighbor electrons is maximal.\footnote{These
TT-states can also be obtained as the limits of the bulk Moore-Read wave functions generalizing the methods of Appendix \ref{appE} to the torus.} Applying $T_2$ we see that $01$ is two-fold degenerate
whereas $0110$ is four-fold degenerate---this gives the six-fold degeneracy of the ground state. By joining strings of $01$ and $0110$ ground states
it follows, using the Su-Schrieffer counting argument, that the domain walls are quasiparticles with charge $\pm e/4$. In
this way one can construct a general state with $k$ quasiholes  and $2n-k$ quasielectrons and show that it has degeneracy $2^{n-1}$
for fixed positions of the excitations. Thus, again the qualitative properties of the state are obtained on the thin torus.
Moreover, the manifest particle-hole symmetry of the Fock space formulation allows the construction of a general state of quasielectrons and quasiholes.

The Moore-Read pfaffian state is not particle-hole symmetric; conjugating the six degenerate states give six orthogonal states, the anti-pfaffian states, that are believed to
describe a different phase of matter \cite{apf1,apf2}. The TT-limits  $01$ and $0110$ of the pfaffian states  are on the other hand particle-hole symmetric and hence
the anti-pfaffian states have the same TT-limit. For small $L_1$, the difference shows up in subleading terms \cite{emiljuha}.

The analysis of the non-abelian states in the TT-limit has recently been generalized to more general parafermionic states \cite{parafermionic}. Again a simple understanding of the quasiparticles as domain walls between degenerate ground states gives the non-trivial degeneracies \cite{haldaneAPS,read06,weunpublished}.

\begin{acknowledgments}
We would like to thank Thors Hans Hansson, Maria Hermanns, Jainendra Jain, Janik Kailasvuori,  Steven Kivelson, Roderich Moessner, Nick Read, Juha Suorsa, Susanne Viefers, and Emma Wikberg for useful discussions.
This work was supported by the Swedish Research Council and by NordForsk.
\end{acknowledgments}

\appendix

\section{One Landau level on torus}\label{appA}

We here give details for a single Landau level on a torus, which we assume has lengths $L_1,L_2$ in the $x$ and
$y$-directions respectively. The complete analysis was given by Haldane; this includes an arbitrariness in a choice of two-dimensional
 lattice \cite{Haldane85, Haldane85PRL}, see also Ref. \cite{fradkin, read}. As we are interested in the mapping to a one-dimensional system, we
restrict ourselves to the corresponding lattice; this allows for an explicit and simple construction.

In Landau gauge, ${\bf A}=By\hat {\bf x}$,
the hamiltonian for a free electron becomes, in units where $\hbar= c/eB=1$,
\be{freeham}
H=\frac 1 {2m} ({\bf p} - \frac e c {\bf A})^2=-\frac 1 {2m}[(\partial_x-iy)^2+\partial_y^2] \ .
\ee
The invariance under continuous spatial translations in the $y$-direction has now been broken to discrete translations, $y\rightarrow y+n2\pi/L_1$,
where $n$ is an integer: The gauge transformation
$\psi \rightarrow e^{i\Lambda ({\bf r})} \psi$, ${\bf A} \rightarrow {\bf A}+\frac c e \grad \Lambda$ must be periodic $e^{i\Lambda(x+L_1)}=e^{i\Lambda(x)}$
and hence translations $y \rightarrow y+ a$, where $a$ is a constant, can be compensated for in $A_x$, and hence in \pref{freeham}, by the gauge transformation
$e^{-i a x}$ only if $a$ is a multiple of the lattice constant $2\pi /L_1$.

The magnetic translation operators $\hat{t}_{\alpha}, \alpha=1,2$, that translate an electron a distance
$L_{\alpha}/N_s$ in the $\alpha$-direction are
\be{t1t2}
\hat{t}_1=e^{(L_1/N_s)\partial _x}\ , \ \ \ \hat{t}_2=e^{(L_2/N_s)(\partial _y +ix)}\ ,
\ee
where $N_s=L_1 L_2/(2 \pi)$ is the number of flux quanta through the surface; the operators obey
\be{alg}
\hat{t}_1\hat{t}_2=e^{2\pi i/N_s}\hat{t}_2\hat{t}_1 \ \ .
\ee
The Landau level preserving "guiding center" coordinates used in the Read-Rezayi state \pref{rr} are defined as:
\be{quiding}
R_{ix}=(2\pi/L_1)(x_i-i\partial_{y_i})\ , \ \  R_{iy}=(2\pi /L_2)i\partial_{x_i} \ \ .
\ee

The states 
\be{psik}
\psi_k&=&e^{-2(k \pi/L_1)^2}\hat{t}_2^k \psi_0\ ,  \ \ \ k=0,1,...,N_s-1 \ , \\
\psi_0&=&\pi^{-1/4}L_1^{-1/2}\sum_n e^{inL_2x} e^{-(y+nL_2)^2/2} \ \ ,
\ee
form a basis of one-particle states in the lowest Landau
level.\footnote{The mapping to the one-dimensional model can be done in any one Landau level. Replacing the lowest Landau level
wave functions $\psi_k$ with the ones in a single higher Landau level in the end only affects the matrix elements $V_{km}$ in \pref{ham}.}
$\psi_k$ is a periodic gaussian located along the line
$y=-2\pi k/L_1$ and is a  $\hat{t}_1$ eigenstate, $\hat{t}_1\psi_k=e^{i2\pi k/N_s}\psi_k$. Letting $c^{\dagger}_k$ create an electron in state $\psi_k$,
$\{c_k,c_m^\dagger \}=\delta_{km}$, maps the Landau level onto a one-dimensional lattice model with lattice constant
$2\pi/ L_1$, see Fig. \ref{mapping}. A basis of many-particle states is given by $|n_0,n_1,\dots n_{N_s-1}\rangle$, where $n_k=0,1$ depending
on whether site $k$ is empty or occupied by an electron; alternatively the state is described by the positions $\{k_1,k_2,...,k_{N_e}\}$ of the particles.

On the cylinder, the single-particle states in Landau level $p$ are
\begin{eqnarray}\label{psicyl}
\psi_{pk}(\mathbf{r})&=&(\sqrt \pi 2^p p! L_1)^{-1/2}e^{-2(\pi k /L_{1})^2} \nonumber \\
&\times& H_p(y+k\frac {2\pi} {L_1})e^{2\pi ikz/L_1}e^{-y^{2}/2}\ \ ,
\end{eqnarray}
where $z=x+iy$ and $H_p$ is the $p:$th Hermite polynomial. Here, the lowest Landau level wave functions, $\psi_{0k}\equiv \psi_k$,  are
obtained by setting $n=0$ in  \pref{psik}.

Consider the electron gas at filling fraction $\nu=p/q$, where $p$ and $q$ are relatively prime integers and the number of
electrons, $N_e=N_s p/q$, is an integer. The operators $T_{\alpha}=\prod_i^{N_e} \hat{t}_{i\alpha}$, (where $\hat{t}_{i\alpha}$ translates electron $i$)
commute with $H$ and, since $T_{\alpha}^{N_s}=1$, the eigenvalues are
$e^{2\pi iK_{\alpha} /N_s}, K_{\alpha}=0,...,N_s-1$. However, $T_1$ and $T_2$ do not commute:
\be{T1T2}
T_1T_2=e^{2\pi i p/q}T_2T_1 \ \ ;
\ee
thus $T_1$ and $T_2^q$ commute and $\{H, T_1, T_2^q\}$ is a maximal set of commuting
operators. $T_2$ changes $K_1$ by $N_e$ and leaves the energy unchanged. Hence, each energy eigenstate is $q$-fold degenerate and we can
choose to characterize it  by the smallest $K_1$. Thus, the energy eigenstates are characterized by a two-dimensional vector $K_\alpha=0,...,N_s/q-1$;
the eigenvalues of  $T_1$ and $T_2^q$ are $e^{2\pi iK_{1} /N_s}$ and $e^{2\pi i q K_{2} /N_s}$ respectively.
The vector $K_\alpha$ corresponds to, but is different from, Haldane's vector  $k_\alpha$ in Ref. \onlinecite{Haldane85PRL}, which characterizes the relative
motion of the electrons only. 

The TT-state that is the ground state at $\nu=p/q=N_e/N_s$ in the TT-limit is a single Slater determinant  $\{k_1,k_2,\dots k_{N_e}\}$ with eigenvalues
\be{}
&&T_1\Psi_{TT}=e^{2\pi i \sum_i k_i /N_s}\Psi_{TT} \nonumber \\
 &&T_2^q\Psi_{TT}=(-1)^{p(N_e-p)}\Psi_{TT} \ \ \  .
\ee
These TT-states are continuous limits of QH hierarchy states,
hence the latter have the same quantum numbers.

The second quantized electron-electron interaction is
\begin{eqnarray}H=\sum_{k_1,
k_2, k_3, k_4=0}^{N_s-1}\!\!\!
V_{k_1k_2 k_3
k_4}c^{\dagger}_{k_1}c^{\dagger}_{k_2}c_{k_3}c_{k_4}\ \ , \label{hamf}\end{eqnarray}
where the matrix elements are
\begin{eqnarray}V_{k_1 k_2 k_3
k_4}=\frac 1 2\int\int d^{2}r_1d^{2}r_2
\psi_{k_1}^*(\mathbf{r}_1)\psi_{k_2}^*(\mathbf{r}_2)\nonumber\\\times V(|\mathbf{r}_1-\mathbf{r}_2|) \psi_{k_3}(\mathbf{r}_2)\psi_{k_4}(\mathbf{r}_1) \ ;
\end{eqnarray}
here both the one-particle states $\psi_k$ and the interaction $V(\mathbf{r})$ are periodic and the integration is over the torus with sides $L_1$ and $L_2$ \cite{yoshioka}.
For the Coulomb interaction, $V(\mathbf{r})=\sum_{s,t}\frac{e^2}{\epsilon |\mathbf{r}+sL_1\hat{x}+tL_2\hat{y}|}$, the matrix elements become 
\begin{eqnarray}
V_{k_1k_2k_3k_4}\!=\!\frac{\delta'_{k_1+k_2,k_3+k_4}}{2L_1L_2}\!\!\!\sum_{(q_1,q_2)\neq (0,0)}\delta'_{k_1-k_4,q_1L_1/2\pi}\nonumber\\\times\frac{2\pi
e^2}{\epsilon q}
e^{-\frac{q^2}{2}-i(k_1-k_3)\frac{q_2L_2}{N_s}},\label{matelem}
\end{eqnarray}
where $\delta'$ is the periodic Kronecker delta function (with period $N_s$)
and $q_i=\frac{2\pi n_i}{L_i}$, $n_i=0,\pm 1, \ldots$. The divergent
$\mathbf{q}=0$ term is excluded  in (\ref{matelem}); it would be cancelled by
adding a positive (neutralizing) background
charge. By taking advantage of the translation invariance and
momentum conservation one can re-write the hamiltonian as in \pref{hamf}:
\be{}
H=\!\!\sum_{n=0}^{N_s-1} \sum_{|m| < k\leq
\frac{N_s} 2}\frac{V_{km}}{\!1\!+\!\delta_{k,\frac{N_s}2}\!\!}
c^\dagger_{n+m}c^\dagger_{n+k}c_{n+k+m}c_n \ ,
\ee
where
\begin{eqnarray}
V_{km}=V_{n+m,n+k,n+m+k,n}-V_{n+m,n+k,n,n+m+k}\nonumber\\+V_{n+k,n+m,n,n+m+k}-V_{n+k,n+m,n+m+k,n} \ .
\end{eqnarray}
It follows that the matrix elements are real and
$V_{km}=V_{k,-m}$, which assures that $H$ is hermitian. Moreover, from the periodicity $V_{km}=V_{k+nN_s,m+n'N_s}$ and the symmetry
$V_{km}=V_{-k,m}$ of (\ref{matelem}) follows that $V_{km}=V_{N_s-k,m}$;  this is  used to express the excitation energies in terms of
 $V_{km}$ in Sec. \ref{hierarchy}.

A general many-body eigenstate of a translation invariant hamiltonian can be separated into a center of mass piece, a relative part and a gaussian factor. Prime examples are the Laughlin wave functions at filling fraction $\nu=1/{t_1}$ \cite{Haldane85}
\begin{eqnarray} \Psi_{\frac 1 {t_1}}=\vartheta^{m/{t_1}+(N_s-{t_1})/2{t_1}}_{({t_1}-N_s)/2}({t_1}\sum_i z_i /L_1 | i {t_1} L_2/L_1)\nonumber\\
\times\prod_{i<j} \vartheta_1 ( (z_i-z_j) /L_1 | i L_2/L_1)^{t_1} e^{-\frac
1 2\sum_i y_i ^2 } \ . \label{laughlinto}
\end{eqnarray}
Here $\vartheta_b^a(z | \tau)=\sum_n e^{i\pi
\tau(n+a)^2+2\pi i (n+a)(z+b)}$ (sum over all integers) are the
Jacobi theta functions and $\vartheta_1 \equiv \vartheta_{1/2}^{1/2}$ is
the standard odd theta function. $m=0,1,\ldots,{t_1}-1$ gives the ${t_1}$
degenerate states that differ by a translation of the center of mass only.   QH hierarchy wave functions are
given for the torus geometry in Ref. \onlinecite{toruswfs}.
The non-abelian Moore-Read state is another example of a state that is known on the torus \cite{pfafftorus}.

\section{Ground states as $L_1 \rightarrow 0$}\label{appB}
We here prove that the relaxation procedure in Sec. \ref{solution} gives the ground state.
The energy $E$ of a state $|n_0n_1\dots n_{N_s-1}\rangle $ with $N_e$ electrons can be written in terms of the interaction between the $N_e(N_e-1)$
different  ordered pairs of electrons. Let $\alpha=1,\dots, N_e$ number the electrons along the circle in positive direction, {\it cf} Fig. \ref{circle}. Consider an ordered
pair of electrons $(\alpha, \alpha+k )$  and let $E_{\alpha,\alpha+k}$ be the interaction energy between these electrons
taken along the path, in the positive direction, from $\alpha$ to $\alpha +k$.\footnote{The electrons are numbered from 1 to $N_e$, hence $\alpha + k$
should be understood as
$( \alpha+k) \, {\rm mod} (N_e)$.} Each ordered pair is obtained exactly ones by letting $k=1,\dots N_e-1$,
hence the energy becomes
\be{energy}
E=  \sum_{k=1}^{N_e-1}\sum_{\alpha=1}^{N_e}E_{\alpha,\alpha+k} \equiv \sum_{k=1}^{N_e-1} E^{(k)}\ \ ,
\ee
where  $E^{(k)}$ is the interaction energy for all pairs that are $k$:th nearest neighbors.\footnote{The two orderings of a pair are, in general,  contained in different $E^{(k)}$.}

The crucial observation is that it is possible to minimize the energies  $E^{(k)}$ separately for an interaction that obeys the
concavity condition \pref{gencond} \cite{hubbard}.
This condition  implies that the interaction energy of one electron
with two other electrons that have fixed positions is minimized if the first electron is as close to the
midpoint between the fixed electrons as possible, {\it ie} if the distances to the two fixed electrons differ by at most one lattice constant,
see Fig. \ref{concavity}.

We will show that the state given by the relaxation procedure in Sec. \ref{solution} minimizes all the energies $E^{(k)}$ and hence it minimizes $E$ and is the ground state.

The energy $E^{(k)}$ is minimized as follows. Let $a^{(k)}_\alpha$ denote the distance between electrons $\alpha$ and $\alpha+k$.
Then $\sum_{\alpha=1}^{N_e}  a^{(k)}_\alpha = kN_s$
(in units of the lattice constant), hence the average distance between two electrons that are $k$:th nearest neighbors is  $(a^{(k)}_\alpha)_{ave}=kN_s/N_e=k/\nu$.
This is a trivial consequence of the periodic boundary conditions on the circle and holds for any state $|n_0n_1\dots n_{N_s-1}\rangle $ with $N_e$ electrons.
Using the concavity condition \pref{gencond} as in Fig. \ref{concavity} it follows that $E^{(k)}$ is minimized if the distances $a^{(k)}_\alpha$ between the electrons
in all the pairs of $k$:th nearest neighbors are as equal as possible, {\it ie} if $a^{(k)}_\alpha$ for each fixed $k$ and $\alpha=1,\dots N_e$ differ by one lattice constant at most.
First we note that if $k=p$ then the average distance is an integer, $(a^{(p)}_\alpha)_{ave}=p/\nu=q$. $E^{(p)}$ is thus minimized if and only if the distance from any electron
to its the $p:$th neighbor is $q$; thus the state is periodic with a unit cell of length $q$ containing $p$ electrons.
Consider now the pairs of $k$:th nearest neighbors. Before moving
the electrons to the closest sites, the separations between the electrons in the pairs are all equal, $a^{(k)}_\alpha|_0=k/\nu$, $\alpha=1,\ldots, N_e$.
Each electron is then moved at most 1/2 lattice constant and hence $|a^{(k)}_\alpha-k/\nu| < 1$; note that this is strictly less than one since two electrons never move 1/2 lattice constant each. It then follows that  $|a^{(k)}_\alpha -a^{(k)}_\beta| < 2$ for any $\alpha, \beta$.  But this difference is by construction an integer,
hence $|a^{(k)}_\alpha -a^{(k)}_\beta| \le 1$.
This proves that the unit cell constructed by the relaxation procedure minimizes $E^{(k)}$, for any $k$, and hence gives the ground state.

We believe the ground state solution is unique, up to the center of mass degeneracy, and have verified this in special cases but have not proven it for general $\nu$.

\section{Quasiparticle charge}\label{appC}

We here show that the
quasiparticles proposed in Sec. \ref{solution} have the expected
charges $e^*=\pm e/q$ at $\nu=p/q$.  Letting $\nu_i = p_i/q_i$, we
find from \pref{iterative} \be{pqiterative} p_i=t_ip_{i-1}+\alpha_i
p_{i-2} \ \ , \ \ \  q_i=t_iq_{i-1}+\alpha_i q_{i-2} \ \ ,
\ee where $\alpha_i=1$ ($-1$) is for the upper (lower) equation
in \pref{iterative}. Consider the quasiparticles at $\nu_{n}$,
{\it ie} in the ground state with unit cell ${\bf C}^{(n)}$. The charge of the insertion
${\bf C}^{(n-1)}$ is determined using the Su-Schrieffer counting
argument. Insert ${\bf C}^{(n-1)}$ in $q_{n}$ well separated
places and remove $q_{n-1}$ unit cells ${\bf C}^{(n)}$. This
preserves the number of sites $N_s$ but changes the number of
electrons by $\delta N_e=q_{n}p_{n-1}-q_{n-1}p_{n}$. Thus the
charge of the insertion ${\bf C}^{(n-1)}$ is $e^*=-e\delta
N_e/q_{n}$. From \pref{pqiterative}, we find
\be{pq2}
q_{i}p_{i-1}-q_{i-1}p_{i}=-\alpha_{i}
(q_{i-1}p_{i-2}-q_{i-2}p_{i-1}) \ \ . \ee
Iterating this we find
\begin{eqnarray}\label{pqfinal}
q_{n}p_{n-1}-q_{n-1}p_{n}
&=&\prod_{i=2}^{n}(-\alpha_i)(q_{1}p_{0}-q_{0}p_{1})\nonumber \\
&=& (-1)^{n}\prod_{i=2}^{n}\alpha_i\ \ \ ,
\end{eqnarray}
where the
last step follows from the explicit forms for $\nu_0$ and $\nu_1$ in
Sec. \ref{solution}. This gives $\delta N_e=\pm 1$ and hence
the insertion ${\bf C}^{(n-1)}$ at $\nu_{n}=p_{n}/q_{n}$ has
charge $e^*=\pm e/q_{n}$. The conjugate $\overline{ {\bf
C}^{(n-1)}}$ obviously has the opposite charge $e^*=\mp e/q_{n}$
since $ {\bf C}^{(n-1)}\overline{ {\bf C}^{(n-1)}}= {\bf C}^{(n)}$
is the unit cell of the ground state and hence is neutral (with
respect to the ground state).

\section{Excitation energy as $L_1\rightarrow 0$ }\label{appD}

We here provide background material for the calculation of the energy of a particle-hole pair in Sec. \ref{solution}. We consider explicitly the
ordering in \pref{pp}, but the arguments generalize immediately to \pref{pp2}.  First we show that the replacement \pref{ppnn},
which creates a nearest neighbor particle-hole pair,  simply amounts to moving one electron one lattice constant in the ground
state unit cell ${\bf C}^{(n)}$.
Using \pref{iterative}, and the definition of $\overline {{\bf C}^{(n-1)}}$, it is straightforward to show that
${\bf C}^{(n-1)} \overline {{\bf C}^{(n-1)}} $ and $\overline {{\bf C}^{(n-1)}} {\bf C}^{(n-1)} $ only differ in the ordering of
${\bf C}^{(n-2)}$ and $\overline {{\bf C}^{(n-2)}}$;  thus the original problem is mapped onto the equivalent  problem at the previous level in the hierarchy.
Iterating this all the way to level one we can use the explicit form of the unit cells ${\bf C}^{(1)}$ and $\overline {{\bf C}^{(1)}}$ to verify that one electron has
moved on lattice constant only.

To prove the induction step, we first consider the case where the last condensation leading to ${\bf C}^{(n)}$ is of the
first type in \pref{iterative}. Using $\overline{ {\bf C}^{(n-1)}}={\bf C}^{(n-1)}_{t_n-1}{\bf C}^{(n-2)} $ we find
\begin{eqnarray}
 {\bf C}^{(n-1)} \overline{ {\bf C}^{(n-1)}}={\bf C}^{(n-1)} {\bf C}^{(n-1)}_{t_n-1}{\bf C}^{(n-2)}
 \nonumber \\ = {\bf C}^{(n-1)}_{t_n-1}{\bf C}^{(n-2)}\overline{ {\bf C}^{(n-2)} }{\bf C}^{(n-2)}
\end{eqnarray}
and
\begin{eqnarray}
 \overline{ {\bf C}^{(n-1)}} {\bf C}^{(n-1)}={\bf C}^{(n-1)}_{t_n-1}{\bf C}^{(n-2)}{\bf C}^{(n-1)}
 \nonumber \\ = {\bf C}^{(n-1)}_{t_n-1}{\bf C}^{(n-2)}{\bf C}^{(n-2)} \overline{ {\bf C}^{(n-2)} } \ ,
\end{eqnarray}
which clearly only differ in the ordering of ${\bf C}^{(n-2)}$ and $\overline {{\bf C}^{(n-2)}}$.
For the second type in \pref{iterative} we identify $\overline{ {\bf C}^{(n-1)}}={\bf C}^{(n-1)}_{t_n-2}\overline {{\bf C}^{(n-2)}} $ and find
\begin{eqnarray}
 {\bf C}^{(n-1)} \overline{ {\bf C}^{(n-1)}}={\bf C}^{(n-1)}_{t_n-1}\overline {{\bf C}^{(n-2)}}
 \nonumber \\ = {\bf C}^{(n-1)}_{t_n-2}{\bf C}^{(n-2)}\overline{ {\bf C}^{(n-2)} }\ \overline{ {\bf C}^{(n-2)} }
\end{eqnarray}
and
\begin{eqnarray}
 \overline{ {\bf C}^{(n-1)}} {\bf C}^{(n-1)}={\bf C}^{(n-1)}_{t_n-2}\overline {{\bf C}^{(n-2)}} {\bf C}^{(n-1)}
 \nonumber \\ = {\bf C}^{(n-1)}_{t_n-2}\overline{ {\bf C}^{(n-2)} } {\bf C}^{(n-2)} \overline{ {\bf C}^{(n-2)} } \  ,
\end{eqnarray}
which also differ only in the ordering of ${\bf C}^{(n-2)}$ and $\overline {{\bf C}^{(n-2)}}$. This concludes the proof.

Note that the creation of a nearest neighbor
pair as in \pref{ppnn} amounts to a translation of a single unit cell assuming periodic boundary conditions on this cell.
The finite distance case \pref{pp} amounts to repeating this for $s$
consecutive unit cells, {\it ie} translating them with periodic boundary conditions on each cell separately (this is actually equivalent to translating all
$s$ cells together with periodic boundary conditions on the set). Since one electron is moved one lattice constant for each unit cell,  $s$ electrons
forming a lattice with lattice constant $q$ are moved rigidly one lattice constant in \pref{pp}.

We now derive the expression \pref{energypair} for the energy of a particle-hole pair. To do so we divide the electrons into three disjoint sets $A,B,C$, where set $A$ consists
of the $s$ electrons that moved under \pref{pp} (these electrons are all in the region between the particle and hole). We now note that the action of \pref{pp} can alternatively
be achieved by translating first all unit cells so that the particle-hole region agrees with the one obtained using \pref{ppnn}, and then  translating all unit cells
outside the particle-hole pair region back to their initial positions. The first step does not change any interactions while the second step moves one electron per unit cell outside of the particle-hole pair region---these electrons form set $B$. Again, the electrons in set $B$ form a regular lattice with lattice constant $q$. Set $C$ consists of the remaining electrons.
From the first construction follows that all the relative distances between the particles in the set $B\cup C$ are unchanged and from the second that the same holds true for
the particles in $A \cup C$; in addition all relative distances within each of the three sets are unchanged.

The energy of a state is the sum over the interaction energies of all pairs of electrons, and the energy of an excitation relative to the ground state is the sum of
the change of all such terms. It follows from the above that a pair energy can change only if one particle is in $A$ and the other is in $B$.
In the ground state the electrons in $A\cup B$ form a lattice with lattice constant $q$; the particle-hole excitation is obtained by translating a string of $s$ adjacent
of these electrons one lattice constant.
This shows which pair energies change; adding the contributions gives \pref{energypair}. 

We now show that the nearest neighbor particle-hole-pair in \pref{ppnn} is the lowest energy excitation
at a given filling fraction $\nu=p/q$. The ground state is the state that minimizes 
the interaction energy between all pairs of $k$:th nearest neighbors $E^{(k)}$ in \pref{energy} for all $k=1,...,N_e-1$.
For an excited state at least one  $E^{(k)}$ must be non-minimal. We divide the
excited states into two categories: the states where $E^{(np)}$, $n=1,2,\ldots, (N_e-1)/p$,  are minimal and those
where they are not. The former are the periodic states with exactly $p$ electrons on any
$q$ consecutive sites. The nearest neighbor particle-hole excitation is obtained by moving one
electron one lattice constant and is hence a non-periodic state. We have seen above that only the distances to the
$np$:th neighbors change, hence only $E^{(np)}$ change for this excitation. Furthermore, the change in
$E^{(np)}$ is  the smallest possible, as implied by the concavity condition (\ref{vcond}) (see also Fig. \ref{concavity}),
thus the nearest neighbor particle-hole pair is the smallest energy non-periodic excitation. The periodic excitations have a much higher energy
since they violate the ground state conditions within each unit cell, thus their energy diverge
with the size of the system.

\section{Laughlin and Jain states as $L_1 \rightarrow 0$}\label{appE}

Here we show that  the $L_1 \rightarrow 0$ limit of the Laughlin and Jain states, and of their quasiparticle excitations,  are the TT-states and
quasiparticles given in Sec. III.  For the ground state at $\nu=1/3$ this was shown in Ref. \onlinecite{Haldane94}.

For simplicity, we give the argument on the cylinder rather than on the torus. Expressing the wave functions in terms of $z$ and $y$,
the projection onto the lowest Landau level is achieved by  moving $y$ to the left and letting $y\rightarrow i\partial _z$, where the derivative should
not act on the factor $e^{-y^2/2}$; this is a straightforward generalization of the method given by  Girvin and Jach \cite{proj}.
Throughout this Appendix the overall factor $e^{-\sum_i y_i^2/2}$ is omitted in all  wave functions as it is not affected
by the projection.

The Jain state at $\nu=p/(2mp+1)$ on the cylinder takes the form \cite{jainrev2}
\be{jaincyl}
\Psi_{\frac p {2mp+1}}= {\rm Det} [\hat \eta_{j}(\hat {\bf r}_i)\gamma _i] \ \ \ ,
\ee
where
\be{gamma}
\gamma_i=\prod_{j\neq i}(\beta_i-\beta_j)^m \ , \ \ \ \ \ \beta_i=e^{2\pi iz_i/L_1} \ \ .
\ee
$ \eta_{i}( {\bf r}_j)$ is a Landau level wave function $\psi_{nk}({\bf r}_j)$ \pref{psicyl} 
\be{nthll}
\psi_{nk} ({\bf r})\propto e^{-2 (\frac {k \pi} {L_1})^2} H_n(y+k\frac {2\pi} {L_1})\beta^k \ \ ,
\ee
and $ \hat \eta_{i}(\hat {\bf r}_j)$ is the
operator obtained from this wave function by letting $y_j\rightarrow i\partial _{z_j}$. The $ \eta_{i}( {\bf r}_j)$ are
chosen as the one-particle states in the $p$ lowest Landau levels.  (Note that $\partial _z$ acts within one term in the determinant;
this corresponds to Jain's second method of projecting onto the lowest Landau level \cite{jainrev2}.)

Consider first the Laughlin states, $\nu=1/(2m+1)$, then $\eta$ are the lowest Landau level wave functions 
\be{eta}
\eta({\bf r})=\psi_{0k} \propto e^{-2 (\frac {k \pi} {L_1})^2} \beta^k \ \ .
\ee
(Only a single Slater determinant of one-electron states survives when $L_1 \rightarrow 0$, hence we can ignore multiplicative
$L_1$-independent constants
in the wave functions.) In this case there is no dependence on $y$ and no need to project onto the lowest Landau level; \pref{jaincyl} of course
gives the Laughlin wave function on the cylinder \cite{thouless}
\be{laughlin}
\Psi_{\frac 1 {2m+1}} = \prod_{i<j} (\beta_i-\beta_j)^{2m+1} \ \ .
\ee
We want to translate this wave function to an occupation number basis, {\it ie} to write it as a sum of Slater determinants of one-electron states. This is
achieved by noting that $\beta^k$ corresponds to the unique  lowest Landau level one-electron state $\psi_{k}$ \pref{psicyl}. Thus,  replacing
\be{replace}
\beta^k \rightarrow   e^{2 (\frac {k \pi} {L_1})^2} \psi_{0k}
\ee
in \pref{jaincyl}, or in this case in \pref{laughlin},  the Slater determinants can be read off;
these states all have the same $K_1=\sum k_i$. Because of the factor in \pref{replace}  the state with the
largest $\sum _i k_i^2$ dominates when $L_1 \rightarrow 0$. This is clearly the state obtained by first choosing
one $k$ as large as possible, then the next as large as possible and so on (all $k_i$ are different because of the Pauli principle).
It is easy to see that the Tao-Thouless state with unit cell $0_{2m}1$ is
contained in \pref{laughlin} and that it is the state which maximizes $\sum_i k_i^2$ for the given $K_1= \sum_i k_i$. Hence, it is the limit
of $\Psi_{\frac 1 {2m+1}}$ as $L_1 \rightarrow 0$. The quasiparticles are discussed below.

We now turn to the Jain states, where $p>1$.  The determinant in \pref{jaincyl} then contains the wave functions, $\psi_{nk}$, $n=0,1,\dots, p-1$
of the $p$ lowest Landau levels.
Projecting $\psi_{nk}({\bf r}_i) \gamma _i$ onto the lowest Landau level, $y\rightarrow i\partial _z$, gives
\be{}
\hat \psi_{nk}(\hat {\bf r}_i) \gamma _i &\propto& e^{-2 (\frac {k \pi} {L_1})^2}H_n(i\partial_{z_i}+k\frac {2\pi} {L_1})\beta_i^k\gamma_i \nonumber  \\
&=&e^{-2 (\frac {k \pi} {L_1})^2}\beta_i^k H_n(i\partial_{z_i})\gamma_i \ \ .
\ee
Defining
\be{alpha}
\alpha_{it}\equiv m\frac {2\pi} {L_1} \sum_{j\neq i}  (\frac {\beta_i} {\beta_i-\beta_j} )^t \ ,
\ee
we have, using $\partial_{z_i} \beta_i=\frac {2\pi i} {L_1} \beta_i$,
\be{}
\partial_{z_i} \gamma_i &=& \alpha_{i1} \gamma_i \  ,  \nonumber \\
\partial_{z_i} \alpha_{it}&=&t ( \alpha_{it}- \alpha_{i,t+1})  \ \ \ .
\ee
Since the Hermite polynomial $H_n$ is an $n$:th grade polynomial in $\partial_{z_i}$
it follows that
\be{herm}
H_n(i\partial_{z_i})\gamma_i &=& m_{in} \gamma_i \ \ ,
\ee
where $m_{in}$ is a polynomial in $\alpha_{it}$. 

Forming the determinant in \pref{jaincyl}, using \pref{herm} and noting that $\prod_i\gamma_i= \prod_ {i\neq j}(\beta_i-\beta_j)^{m}=\prod_ {i<j}(\beta_i-\beta_j)^{2m}
\equiv \Psi_{\frac 1 {2m}}$  is the $\nu=1/2m$ bosonic Laughlin state, one finds that \pref{jaincyl} can be written as
\be{psibeta}
\Psi_{\frac p {2mp+1}} \propto {\rm Det} [M_{\{nk\}i}]\Psi_{\frac 1 {2m}} \ ,
\ee
where
\be{matrix}
M_{\{nk\}i}=\beta_i^km_{in} \  .
\ee
Here,  $\{nk\}$ denotes the first index of the matrix $M$: $n=0,1\dots p-1$ numbers the Landau levels and $k=0,1,\dots N-1$ numbers the
states in each level. The second index numbers the electrons,  $i=1,2,\dots pN$.

We have expressed $\Psi_{\frac p {2mp+1}}$ in terms of $\beta$ and can now
translate it to the occupation number basis using \pref{replace}. Just as above, the leading order term will be the one that maximizes
$\sum_i k_i^2$. One finds that this is obtained by taking the state with maximal $\sum_i k_i^2$ in $\Psi_{\frac 1 {2m}}$---this is the TT-state with
unit cell $0_{2m-1}1$. An element $\beta_i^k$ in $M$ acts as an operator on the occupation number states in $\Psi_{\frac 1 {2m}}$, translating
particle $i$ to the right $k$ steps, {\it ie} increasing its momentum by $k$. The elements $\beta^k_i m_{in}$, $n=1,2,\dots$ also move the $i$:th
particle $k$ steps to the right just as $\beta^k_i$ alone does. This is because $m_{in}$ is a polynomial in $\alpha_{it}$ and it can be seen that such terms
do not change the highest power of $\beta_i$ when acting on $\Psi_{\frac 1 {2m}}$, {\it cf}  \pref{alpha}. Considering the action of a term in the determinant
it follows that it will move groups of $p$ electrons $k$ steps to the right, where $k=0,1,\dots, N-1$. We see that $\sum_i k_i^2$ is maximized if the $p$ leftmost electrons,
{\it ie } the ones with smallest $k_i$, are not
moved, then, moving to the right, the next $p$ electrons are moved one step to the right and so on until finally the rightmost set of $p$ electrons,
the ones with the largest $k_i$,  are moved $N-1$ steps to the right.
This gives the TT-state with unit cell $0_{2m}1(0_{2m-1}1)_{p-1}$.

To clarify the limit, we consider  $\nu=2/5$ explicitly. In this case, $p=2$, $m=1$, and thus there are $2N$ particles---$N$ in each of the two effective Landau levels. 
From (\ref{psibeta}) and (\ref{matrix}) it follows that we can write the wave function as 

\be{25beta}\!\!\! \!\Psi_{\frac 2 5}\!=\!\!\left| \! \begin{array}{cccc} 
\beta_{1}^{N-1} m_{1,1} & \beta_{2}^{N-1} m_{2,1} &\! \ldots \!& \beta_{2N}^{N-1} m_{2N,1} \\ 
\beta_{1}^{N-2} m_{1,1} & \beta_{2}^{N-2} m_{2,1} &\! \ldots\! & \beta_{2N}^{N-2} m_{2N,1} \\ 
\vdots & \vdots &\! \ddots\! & \vdots \\
 m_{1,1} & m_{2,1} &\! \ldots \!& m_{2N,1} \\ 
\beta_{1}^{N-1} m_{1,0} & \beta_{2}^{N-1} m_{2,0} &\! \ldots \!& \beta_{2N}^{N-1} m_{2N,0} \\ 
\vdots & \vdots &\! \ddots \!& \vdots \\
m_{1,0} & m_{2,0} &\! \ldots\! & m_{2N,0} 
\end{array}\! \right|\! \Psi_{\frac 1 {2}}\!.
\ee
To maximize $\sum_i k_i^2$  we pick one $k_i$, {\it ie} one of the powers of $\beta_i$, as large as possible in the expansion of (\ref{25beta}). This is achieved if and only if the power of $\beta_i$ is maximized both in the determinant and in $\Psi_{\frac 1 {2 }}$. 
The highest power in the determinant is $N-1$ and in $\Psi_{\frac 1 {2}}$ it is $2(2N-1)$, thus the maximal $k_i$ is $5N-3$. Next, we need to find the second highest power. Again the power from the determinant is $N-1$ (note that a given term in the determinant can include factors such as $\beta_{1}^{N-1}\beta_{2}^{N-1}$) while it is $2(2N-2)$ from the Jastrow factors $\Psi_{\frac 1 {2 }}$, thus the total power is $5N-5$. Next one finds that the determinant can only contribute a power $N-2$ and that the highest power coming from  $\Psi_{\frac 1 {2 }}$ is $2(2N-3)$. Continuing this one finds that the maximal $k_i$ in the determinant decreases by one for every second particle while it always decreases by two in $\Psi_{\frac 1 {2 }}$. This gives the unit cell $00101$. From this example, the general procedure given above should be clear.

It is now straightforward to generalize the analysis above to quasiparticle states. In Jain's approach, these are constructed
by replacing one of the one-electron states in \pref{jaincyl} by a state in a higher empty Landau level. Assuming the levels involved are the highest
occupied and the lowest empty ones, $\psi_{p-1,k} \rightarrow \psi_{p,q}$,  this amounts to replacing $\beta_i^k m_{i,p-1}$ by $\beta_i^q m_{i,p}$
(for all $i$ and fixed $k,q$) in the matrix $M$ in \pref{matrix}.
This has the effect of moving $0_{2m-1}1$ from one place to another in the TT-ground state, {\it ie} of creating a particle-hole pair of excitations with charge
$\pm e/(2m+1)$, {\it cf}  Sec. III B. 
The position of the excitation depends on $k$ and $q$, and in particular one of them can be put at the end of the cylinder.

\section{Laughlin state on cylinder}\label{appF}

The Laughlin state is the exact and unique ground state
for a certain short-range interaction and there is a gap to all excitations \cite{haldanebook,trugman,pokrovsky}. The
proofs were formulated using spherical or disk geometry.  The essential ingredient is, however, only
the short-distance behavior of the wave functions when two electrons approach each-other---thus
it may be regarded as obvious that the result holds also on the cylinder (or torus) for arbitrary circumference $L_1$; this
result is in fact implicit in Ref. \onlinecite{Haldane94}.
There has, however,  been some controversy on this point and as the result is important in establishing that
the state develops continuously without a phase transition from the TT-state for small $L_1$  to the
two-dimensional bulk Laughlin state as $L_1\rightarrow \infty$, we here explicitly show
 that the Laughlin state
is the exact and, up to rigid translations,  unique  ground state on a cylinder for a short range interaction
and that there is a gap to all excitations for any $L_1$. Our argument follows closely that of Trugman and Kivelson \cite{trugman}.

A fermionic many-particle state in the lowest Landau level is a sum of Slater determinants
\begin{eqnarray}
\Psi(\{\mathbf{r}_i\})\!&=&\!\sum_{\{k_i\}}a_{\{k_i\}}\det_{i,j}(\psi_{k_i}(\mathbf{r}_j)) \nonumber \\
&=& f(\{\beta_i\})e^{-\sum_jy_j^2/2} \ \ ,
\end{eqnarray}
where $\psi_k$ are lowest Landau level single particle states on the cylinder \pref{psicyl} and $\beta_i=e^{2\pi iz_i/L_1}$.
The range $\Delta k$ of the momenta $k_i$ restricts the system in space and thus depends
on the number of particles and the filling factor $\nu=p/q$.
The filling factor is the number of electrons divided by the number of sites,
$\nu=N_e/(\Delta k+1)$. Because of the $q$-fold translational degeneracy, $\Delta k =\Delta \tilde k -(q-1)$,
where $\Delta \tilde k$ is the range of momenta in $f$.
$f$ is an
anti-symmetric polynomial in $\beta_i$, hence it contains the Jastrow factor $J=\prod_{i<j}(\beta_i-\beta_j)$.
The problem of
minimizing the energy of a specific electron-electron interaction in
the lowest Landau level is thus reduced to finding the
polynomial $f$ that gives the lowest energy given the above constraints. (This is in complete analogy with the problem in the
plane where the polynomial is instead in terms of $z_i$.)

Consider now the repulsive interaction
\begin{equation}
V(\mathbf r)=\sum_{s=0}^{\infty}V_s(\mathbf r)=\sum_{s=0}^{\infty}
c_s b^{2s}\nabla^{2s}\delta_p(\mathbf{r})\ \ ,\label{cylint}
\end{equation}
where $b$ is the range of the interaction, $c_s$ are positive constants, and
$\delta_p(x,y)=\sum_n\delta(x+nL_1,y)$ is the periodic
delta-function. The  leading term, $\expect{V_0}$, is identically zero for any
fermionic state. By assuming $b\rightarrow 0$ only the leading non-vanishing term in \pref{cylint} contributes to the energy
$E=\expect{V(\mathbf r)}$.

The expectation value of $V_s$ is
\begin{equation}
\expect{V_s}=4c_s b^{2s}\sum_{i< j} \prod_{k\neq i,j} \int d^2\mathbf{r}_k d^2 Z d^2 z \delta^2_p (z) \partial^s {\bar \partial}^s(\Psi^*\Psi),\label{vs}
\end{equation}
where we have integrated by parts \footnote{The boundary terms vanish as is readily seen by considering the $y_{ij}$-integrals.},
and changed to relative coordinates for the pair that is affected by the interaction, $z=z_i-z_j,Z=(z_i+z_j)/2$, $\partial =\partial/\partial z$.
Performing the derivatives in the integrand gives
\be{expand}
 \partial^s {\bar \partial}^s(\Psi ^* \Psi)  =e^{-\sum_{k\neq i,j}y_k^2/2} e^{(z-\bar z)^2/16}e^{(Z-\bar Z)^2/4} \ \nonumber \\
 \times [\partial ^s f {\bar \partial} ^s {\bar f}
 + \sum_{r,t=0, r+t \neq 2s}^{s} \alpha_{rt}  \partial ^r f{\bar \partial} ^t {\bar f}] \ \ ,
\ee
for some $\alpha_{rt}$.
For $s=1$, only the first term  contributes to $\expect{V_{1}}$ since $f(0)=0$, and 
this term is non-negative  $\expect{V_{1}}\ge 0$.  $\expect{V_{1}} = 0$ if and only if  $\partial f|_{z=0}=0$; if this is the case
then only the first term in \pref{expand} for $s=2$ contributes to  $\expect{V_{2}}$ and
$\expect{V_{2}}\ge 0$, the equality sign holding if and only if $\partial  f|_0=\partial f ^2|_0=0$. This continues to
any order in $s$ and we have that $\expect{V_{t}}=0$ for $t=1,2,\dots s$, and $\expect {V_{s+1}}\ge 0$,  if and only if $\partial ^tf |_0=0$ for $t=1,2,\dots s$. This means that $f$ vanishes
at least as fast as $z^{s+1}$ when $z \rightarrow 0$, and since this holds for any $z_i-z_j=z$ and $f$ is an antisymmetric polynomial in $\beta_i$,
it follows that $f$ contains at least $s+1$ powers of the Jastrow factor: $f=J^{s+1}\tilde f$, where $\tilde f$ is a new polynomial in $\beta_i$.

Consider the truncated interaction $V^{(s)}(\mathbf r)=\sum_{t=0}^{s}V_t(\mathbf r)$. We have found that a state has vanishing energy if and only if
$f=J^{s+1}\tilde f$. These are the ground states: All other states have positive energy since, as $b \rightarrow 0$, only the first non-zero
$\expect {V_t}$ contributes and this is non-negative according to the discussion above.

For $\tilde f=1$, $f$ is the Laughlin state at filling factor $\nu=1/(s+1)$ which thus is a ground state; it is unique since any non-constant polynomial $\tilde f$ decreases the
filling factor.  This proves the proposition.

On the torus, an analogous argument implies that the zero-energy
states must contain a factor $\prod_{i<j} \vartheta_1 ( (z_i-z_j) /L_1
| i L_2/L_1)^{s+1}$, which together with the boundary conditions specify
the Laughlin state on the torus (\ref{laughlinto}) uniquely up to
the $s+1-$fold center of mass degeneracy.

\section{Exact solution at $\nu=1/2$}\label{appG}

Here we give details of the exact diagonalization of the short-range
hamiltonian \pref{shortham}\cite{bk1}.

The crucial part in (\ref{shortham}) is the hopping
term $V_{21}$. However, we begin by considering the
electrostatic part
$H_{sr}|_{V_{21}=0}$, which has eigenstates with fixed charges,
$|n_0n_1...n_{N_s-1}\rangle$. The energy of such a state is
 \beqa \label{elstat2}
E_0=\alpha(\frac {N_s} 2 + n_{111} + n_{000} )\ \ ,
\eeqa
where, $n_{111} \, (n_{000})$ is the number of
3-strings\footnote{A string of length $k\ge 3$ is counted as $k-2$
strings.}, {\it ie}
strings consisting of three nearby electrons (holes) in
$n_0n_1...n_{N_s-1}$. Thus there is a degenerate ground state manifold
${\cal H}_0$ consisting of all states where at most two electrons or
two holes are next to each other. The energy $E_0$ follows
by writing the right hand side of \pref{elstat2} in terms of  $\hat{n}_i$ and comparing to \pref{shortham}. 

To proceed we define a subspace
${\cal H}^\prime$ of the full Hilbert space by requiring each pair
of sites $(2p,2p+1)$ to have charge one (acting with $T_2$ gives an
equivalent  grouping of the sites  $(2p-1,2p)$ and a
corresponding subspace $\cal H ^\prime _T$). As we will demonstrate below, $H_{sr}$
preserves the subspace $\cal H^\prime$ and can be exactly diagonalized in this subspace
giving non-interacting neutral fermions. Note that ${\cal
  H}^\prime\subset{\cal H}_0$, thus any state in ${\cal H}^\prime$ has the lowest possible
  electrostatic energy.  As argued in the main text and in Ref. \onlinecite{bk1}, ${\cal H}^\prime$ also contains
the low-energy sector of the hopping term $V_{21}$ and thus of the hamiltonian $H_{sr}$ (\ref{shortham}).

There are two possible states for a pair of sites in $\cal H ^\prime$:
\begin{equation}
|\downarrow \rangle \equiv |01\rangle \ , \ \ |\uparrow \rangle
\equiv |10\rangle \ \ ;
\end{equation}
introducing the spin operators
\begin{equation}s^+_p=c^\dagger_{2p}c_{2p+1}\ ,\ \ \
s^-_p=c^\dagger_{2p+1}c_{2p}\ \ , \label{spinop}
\end{equation}
where $p=0,1,\dots N_e-1$,
we have
\begin{equation}
s^+|\downarrow \rangle = |\uparrow \rangle \ , \ \ \  s^-|\uparrow \rangle = |\downarrow \rangle \ \ .
\end{equation}
On states in $\cal H ^\prime$,
$s^+, \,
s^-$ describe hard core bosons---they commute on
different sites but obey anti-commutation relations on the same site
\begin{eqnarray}
[s^{-}_{i}, s^{+}_{j}]=[s^{-}_{i}, s^{-}_{j}]=[s^{+}_{i},
s^{+}_{j}]=0\ , \ \ \ i\neq j \nonumber\\ \{s^{-}_{i}, s^{+}_{i}\}=1,
\ \ \{s^{-}_{i}, s^{-}_{i}\}=\{s^{+}_{i}, s^{+}_{i}\}=0 \ \ ,
\end{eqnarray} and $H_{sr}$ is the nearest neighbor spin 1/2 $xy$-chain, 
\begin{equation}
H_{sr}=\alpha N_e+V_{21}
\sum_{p=0}^{N_e-1} [s^{+}_{p+1} s^{-}_{p}+s^{-}_{p+1}
s^{+}_{p}]\ ,
\end{equation}
where $s^{\pm}_{N_e}\equiv s^{\pm}_0$ and $V_{21}>0$.
We diagonalize $H_{sr}$ following the standard procedure, see {\it eg} Ref. \onlinecite{fradkin, schulz}.
Expressing the (hard core) bosons in terms
of fermions $d$ using the Jordan-Wigner transformation,
\begin{equation}
s^-_p=K_p d_p,\ \ \ K_p=e^{i\pi
\sum_{j=0}^{p-1}d^\dagger_j d_j} \ ,
\end{equation} the hamiltonian
becomes that of free fermions,
\begin{equation}
H_{sr}=\alpha N_e+V_{21}\sum_{p=0}^{N_e-2}
[d^\dagger_{p+1} d_{p} +d^\dagger_{0}
K_{N_e-1}d_{N_e-1}+hc]\ ,
\end{equation}
when restricted to $\cal H ^\prime$. The explicit $K$-factor can be ignored for $N_e \rightarrow \infty$ \cite{Lieb}; 
however, $H_{sr}$ can be diagonalized exactly also for finite $N_e$ including this $K$-factor by noting that, in this term, $K_{N_e-1}=(-)^{N_\uparrow +1}$, 
where $N_\uparrow$ is the number of up spins in the state. Thus 
\begin{equation}
H_{sr}=\alpha N_e+V_{21}\sum_{p=0}^{N_e-1}
[d^\dagger_{p+1} d_{p} +d^\dagger_p d_{p+1}]\ ,
\end{equation}
where $d_{N_e}\equiv (-)^{N_\uparrow+1}d_0$. Since $N_\uparrow$ is conserved by $H_{sr}$, we can diagonalize $H_{sr}$
by the Fourier transformation
\begin{eqnarray}
 d_{j}&=&\frac{1}{\sqrt{N_e}}\sum_k e^{ijk}\tilde{d}_k \ , \nonumber \\
 \tilde{d}_{k}&=&\frac{1}{\sqrt{N_e}}\sum_{j=0}^{N_e-1} e^{-ijk}{d}_j \ ,
\end{eqnarray}
where for $N_\uparrow$ odd
\begin{eqnarray}\label{kvalues}
k \frac {N_e} {2\pi}  &=& 0,\pm 1,\pm 2, \dots \pm (\frac {N_e}  2 -1), \frac {N_e} 2 \ \ \  {\rm for }  \ N_e \ {\rm even} \nonumber \\
k \frac {N_e} {2\pi}  &=& 0,\pm  1\pm 2 , \dots \pm (\frac {N_e} 2 -\frac 1 2 ) \ \ \  {\rm for }  \ N_e \ {\rm odd} \, ;
\end{eqnarray}
whereas for $N_\uparrow$ even, $k$ is shifted by $-\pi/N_e$: 
\begin{equation}\label{shiftedk}
k \frac {N_e} {2\pi} \rightarrow k \frac {N_e} {2\pi} -\frac 1 2 \ \ .
\end{equation}
Note that this implies that  different sets of momenta $k$ are used in states obtained when acting with an even or odd number of operators 
$\tilde d^\dagger$ on the Fock vacuum
\begin{equation}\label{dstates}
\prod_{j=1}^{N_\uparrow}\tilde d^\dagger _{k_j}|010101\dots\rangle \  \  .
\end{equation}
It can be shown  that \pref{dstates} are eigenstates of $T_1$ and $T_2^2$ with eigenvalues
\begin{eqnarray}\label{qnumber}
K_1 &=& -N_\uparrow +N_e^2\nonumber \\ 
K_2 &=& -\frac {N_e} {2 \pi} \sum_{j=1}^{N_\uparrow} k_j+\frac {N_e(N_e-1)} 2  \ \  .
\end{eqnarray}

The hamiltonian becomes
\begin{equation}
H_{sr}=\alpha N_e+2V_{21}\sum_{k}\cos{k} \ \tilde{d}^\dagger_{k}
\tilde{d}_{k} \ .
\end{equation} The ground state is obtained by filling the lowest energy states respecting the
condition that an odd (even) number of states must be filled when $k$ takes the values in \pref{kvalues} (is shifted by $-\pi/N_e$).
For $N_e$ even, a unique ground state is obtained by filling all the negative energy states  
\begin{equation}
\ket{GS}=\prod_{|k|>\frac \pi 2}
  \tilde{d}^\dagger_{k}\ket{010101\ldots}\ .
\end{equation}
For $N_e$ odd, the ground state is obtained by filling all the negative states except for the highest, or alternatively all the negative states and the lowest positive state. This leads to a four-fold degenerate ground state (in $\cal H ^\prime $). This degeneracy corresponds to reflections of the Fermi sea in horizontal and vertical lines that give new Fermi seas for odd $N_e$ but not for even $N_e$, see Figs. \ref{boxes},  \ref{phasediagrams}.

This solves the problem in  $\cal H ^\prime $ and, by action of
$T_2$,  in $\cal H ^\prime _T$.  When $N_e \rightarrow \infty$, the ground state  has energy
\begin{equation} E/N_e=\alpha - \frac {2V_{21}}
{\pi}
\end{equation} 
per electron. The excitations are neutral
particle and/or hole excitations out of this Fermi sea. They have, according to \pref{spinop}, a natural
interpretation in terms of dipoles and in the limit $N_e\rightarrow
\infty$ the excitations become gapless. All of the electric charge
sits in the Fock vacuum, $\ket{010101\ldots}$, while the
quasi-particles $\tilde d_k$ that build up the ground state are
neutral.

A straightforward calculation shows that
$\langle c^\dagger_m c_n\rangle=\half \delta_{mn}$, when $N_e\rightarrow \infty$ \footnote{In the lattice description,
the amplitude of the (period two) charge density wave
is $1/N_s=1/(2N_e)$ when $N_e$ is odd and zero for even $N_e$.},
hence
\be{homo}
\langle \rho(\mathbf{r}) \rangle &=& \sum_{m,n}\psi^{*}_{m}(\mathbf{r})\psi_{n}(\mathbf{r})\langle c_{m}^{\dag}c_{n}\rangle \nonumber \\
&=& \frac 1 2 \sum_{m}\psi^{*}_{m}(\mathbf{r})\psi_{m}(\mathbf{r}) \
\ \ . \ee Thus this state is homogenous in occupation space (this is not the case for the TT-states), and hence approximately homogeneous in real space for any $L_1$,
becoming completely homogeneous when $L_1 \rightarrow \infty$ (as is the case also for the TT-states).


\begin{thebibliography} {999}

{\footnotesize


\bibitem{vonk} K.v. Klitzing, G. Dorda and M. Pepper, Phys. Rev. Lett. {\bf 45}, 494 (1980).


\bibitem{tsui}  D.C. Tsui, H.L. St\"ormer and A.C. Gossard, Phys. Rev. Lett. {\bf 48}, 1599 (1982).

\bibitem{stat1} B.I. Halperin,  Phys. Rev. Lett. {\bf 52}, 1583 (1984).

\bibitem{stat2} D.P. Arovas, J.R. Schrieffer, and F. Wilczek, Phys. Rev. Lett. {\bf 53}, 722 (1984).

\bibitem{it1} R.B. Laughlin, Phys. Rev. B. {\bf 23}, 5632 (1981).

\bibitem{it2} B.I. Halperin, Phys. Rev. B. {\bf 25}, 2185 (1982).

\bibitem{Laughlin83} R.B. Laughlin,  Phys. Rev. Lett. {\bf 50}, 1395 (1983); R.B. Laughlin in {\it The Quantum Hall Effect},
eds. R.E. Prange, and S.M. Girvin, (Springer-Verlag, New York, 1990).

\bibitem{jiang} H.W. Jiang, H.L. St\"ormer, D.C. Tsui, L.N. Pfeiffer, and K.W. West, Phys. Rev. B {\bf 40}, R12013 (1989).

\bibitem{willett90} R.L. Willett, M.A. Paalanen, R.R. Ruel, K.W. West, L.N. Pfeiffer, and D.J. Bishop,   Phys. Rev. Lett. {\bf 65}, 112 (1990).

\bibitem{wang} J.K. Wang, and V.J. Goldman, Phys. Rev. B {\bf 45}, 13479 (1992).

\bibitem{striped1}  M.P. Lilly, K.B. Cooper, J.P. Eisenstein, L.N. Pfeiffer, and K. W. West, Phys. Rev. Lett. {\bf 82}, 394 (1999).

\bibitem{striped2} R.R. Du, D.C. Tsui, H.L. St\"ormer, L.N. Pfeiffer, K.W. Baldwin, and K.W. West, Solid State Comm. {\bf  109}, 389 (1999).

\bibitem{mr} G. Moore, and N. Read, Nucl. Phys. B {\bf 360}, 362 (1991).

\bibitem{nayak} C. Nayak, and F. Wilczek, Nucl. Phys. B {\bf 479}, 529 (1996).

\bibitem{readrezayi} N. Read, and E.H. Rezayi,  Phys. Rev. B. {\bf 54}, 16864  (1996).

\bibitem{hierarchyHaldane} F.D.M. Haldane, Phys. Rev. Lett. {\bf 51}, 605
(1983).

\bibitem{hierarchyLaughlin} R.B. Laughlin, Surf. Sci. {\bf 141}, 11 (1984).

\bibitem{hierarchyHalperin} B.I. Halperin, Phys. Rev. Lett. {\bf 52}, 1583, 2390(E)
(1984).

\bibitem{jain89} J.K. Jain,  Phys. Rev. Lett. {\bf 63}, 199 (1989).

\bibitem{jainrev1} J.K. Jain in {\it Perspectives in Quantum Hall Effects}, eds. S. Das Sarma, and A. Pinczuk, (John Wiley \& Sons, New York, 1996).

\bibitem{jainrev2}  J.K. Jain, and R.K. Kamilla, in {\it Composite Fermions},
ed. O. Heinonen, (World Scientific, Singapore, 1998).

\bibitem{jainbook} J.K. Jain,  {\it Composite fermions}, (Cambridge University Press, 2007).

\bibitem{zhang} S.C. Zhang, T.H. Hansson, and S.A. Kivelson,  Phys. Rev.
Lett. {\bf 62}, 82 (1989).

\bibitem{read89} N. Read,  Phys. Rev. Lett. {\bf 62}, 86 (1989).

\bibitem{lopez} A. Lopez, and E. Fradkin,  Phys.
  Rev. B. {\bf 44}, 5246 (1991).

\bibitem{ballistic1} W. Kang, H.L. St\"ormer, L.N. Pfeiffer, K.W. Baldwin,
and K.W. West, Phys. Rev. Lett. {\bf 71}, 3850 (1993).

\bibitem{ballistic2} V.J. Goldman, B. Su,
and J.K. Jain, Phys. Rev. Lett. {\bf 72}, 2065 (1994).

\bibitem{ballistic3} J.H. Smet, D. Weiss, R.H.
Blick, G. L\"{u}tjering, K. von Klitzing, R. Fleischmann, R.
Ketzmerick, T. Geisel, and G. Weimann, Phys. Rev. Lett. {\bf 77},
2272  (1996).

\bibitem{hlr} B.I. Halperin, P.A. Lee, and N. Read, Phys. Rev. B {\bf 47},
7312 (1993).

\bibitem{kalmeyer} V. Kalmeyer and S.-C. Zhang,  Phys.
  Rev. B. {\bf 46}, 9889 (1992).

\bibitem{Willettreview} R.L. Willett, in {\it Composite Fermions},
ed. O. Heinonen, (World Scientific, Singapore, 1998).

\bibitem{Dyakonov} M.I. Dyakonov, in
{\it Recent Trends in Theory of Physical Phenomena in High Magnetic Fields}, eds. I.D. Vagner, P. Wyder, and T. Maniv, (Kluwer, 2003).

\bibitem{read90} N. Read, Phys. Rev. Lett. {\bf 65}, 1502 (1990).

\bibitem{blokwen} B. Blok, and  X.G. Wen, Phys. Rev. B {\bf 43}, 8337 (1991).


\bibitem{bk1} E.J. Bergholtz, and A. Karlhede, Phys. Rev. Lett. {\bf 94}, 26802
(2005).

\bibitem{bk2} E.J. Bergholtz, and A. Karlhede, J. Stat. Mech. (2006) L04001.

\bibitem{we06} E.J. Bergholtz, J. Kailasvuori, E. Wikberg, T.H. Hansson, and A. Karlhede, Phys. Rev. B {\bf 74}, 081308(R) (2006).

\bibitem{hierarchy} E.J. Bergholtz, T.H. Hansson, M. Hermanns, and A. Karlhede, Phys. Rev. Lett. {\bf 99}, 256803 (2007).

\bibitem{conformal} E.J. Bergholtz, M. Hermanns, T.H. Hansson,  A. Karlhede and S.F. Viefers, arXiv:0712.3848 (2007).

\bibitem{Tao83} R. Tao, and D.J. Thouless, Phys. Rev. B {\bf 28}, 1142 (1983).

\bibitem{hans} T.H. Hansson, C.C. Chang, J.K.  Jain,  and S.F. Viefers,  Phys. Rev. Lett. {\bf 98} 076801  (2007).

\bibitem{rr} E.H. Rezayi, and N. Read,  Phys. Rev. Lett. {\bf 72}, 900  (1994).

\bibitem{haldaneAPS} F.D.M. Haldane, Talk at the APS March meeting (2006).

\bibitem{seidel06} A. Seidel, and D.-H. Lee, Phys. Rev. Lett. {\bf 97}, 056804 (2006).

\bibitem{read06} N. Read, Phys. Rev. B {\bf 73}, 245334 (2006).

\bibitem{weunpublished} E. Ardonne, E.J. Bergholtz, J. Kailasvuori, and E. Wikberg, arXiv:0802.0675 (2008). 

\bibitem{haldanejack} B. A. Bernevig, and F.D.M. Haldane, arXiv:0707.3637 (2007).

\bibitem{anderson} P.W. Anderson,  Phys. Rev. B {\bf 28}, 2264 (1983).

\bibitem{su1984} W.P. Su,  Phys. Rev. B {\bf 30}, 1069 (1984).

\bibitem{su2} W.P. Su,  Phys. Rev. B {\bf 32}, 2617 (1985).

\bibitem{Haldane94}  E.H. Rezayi, and F.D.M. Haldane, Phys. Rev. B {\bf 50}, 17199 (1994).

\bibitem{chui1985} S.T. Chui, Phys. Rev. B {\bf 32}, 8438 (1985).

\bibitem{chui86} S.T. Chui, Phys. Rev. Lett. {\bf 56}, 2395  (1986).

\bibitem{heiselberg} H. Heiselberg, Phys. Rev. A {\bf 73}, 013628 (2006).


\bibitem{readearly} N. Read, Semicond. Sci. Technol. {\bf 9}, 1859 (1994).

\bibitem{read.5} N. Read, Phys. Rev. B {\bf 58}, 16262 (1998).

\bibitem{ms} R. Shankar, and G. Murthy, Phys. Rev. Lett. {\bf 79}, 4437  (1997); Rev. Mod. Phys. {\bf 75}, 1101 (2003).

\bibitem{pasq} V. Pasquier and F.D.M. Haldane,  Nuclear Physics
B {\bf 516}, 719 (1998).

\bibitem{dhlee} D.-H. Lee, Phys. Rev. Lett. {\bf 80}, 4745
(1998), Erratum  {\bf 82}, 2416 (1999).

\bibitem{stern} A. Stern, B.I. Halperin, F. von Oppen and S.H. Simon, Phys. Rev. B {\bf 59}, 12547 (1999)

\bibitem{rev1} {\it Composite Fermions}, ed. O. Heinonen, (World Scientific, Singapore, 1998). 

\bibitem{haxton} J.N. Ginocchio and W.C. Haxton, Phys. Rev. Lett. {\bf 77}, 1568 (1996).

\bibitem{wojs} A. Wojs, K.-S. Yi, and J.J. Quinn,  Phys. Rev. B {\bf 69}, 205322 (2004).


\bibitem{greiter94} M. Greiter, Phys. Lett. {\bf B 336}, 48 (1994).


\bibitem{pan} W. Pan, H.L. St\"ormer, D.C. Tsui, L.N. Pfeiffer, K.W. Baldwin, and K.W. West,  Phys. Rev. Lett. {\bf 90}, 016801 (2003).


\bibitem{Haldane85PRL} F.D.M. Haldane, Phys. Rev. Lett. {\bf 55}, 2095 (1985).

\bibitem{hubbard} J. Hubbard, Phys. Rev. B {\bf 17}, 494 (1978).

\bibitem{wen} X.-G. Wen,  and Q. Niu, Phys. Rev. B {\bf 41} 9377 (1990).

\bibitem{wenreview}  X.-G. Wen,  Advances in Physics {\bf 44}, 405 (1995).

\bibitem{Schrieffer} W.P. Su, and J.R. Schrieffer, Phys. Rev. Lett. {\bf 46}, 738 (1981).

\bibitem{mani} R.G. Mani, and K.  Z.  von Klitzing, Phys. B {\bf 100}, 635 (1996).

\bibitem{goerbig} M.O. Goerbig, P. Lederer, and C. Morais Smith,  Europhysics Letters {\bf 68}, 72 (2004).

\bibitem{global} S.A. Kivelson, D.-H. Lee, and S.-C. Zhang,  Phys. Rev. B {\bf 46}, 2223 (1992).

\bibitem{lutken} C.A.  L\"utken, and  G.G. Ross, Phys. Rev. B {\bf 45}, 11837 (1992);
Phys. Rev. B {\bf 48}, 2500 (1993).

\bibitem{goldman} V.J.  Goldman, and M. Shayegan, Surf. Sci. {\bf 229}, 10 (1990).

\bibitem{emergent} S. He, X.C. Xie, and F.C. Zhang, Phys. Rev. Lett. 68, 3460 (1992).

\bibitem{haldanebook} F.D.M. Haldane in {\it The Quantum Hall Effect},
eds. R.E. Prange, and S.M. Girvin, (Springer-Verlag, New York, 1990).

\bibitem{trugman} S.A. Trugman and S.A. Kivelson,  Phys. Rev. B {\bf 31}, 5280 (1985).

\bibitem{pokrovsky} V.L. Pokrovsky and A.L. Talapov, J. Phys. C 18, L691 (1985).

\bibitem{Lee05} A. Seidel, H. Fu, D.-H. Lee, J.M. Leinaas, and J. Moore, Phys. Rev. Lett. 95, 266405 (2005).

\bibitem{Jansen} S. Jansen, E.H. Lieb,  and R. Seiler, arXiv:0711.1467  (2007).

\bibitem{JainGoldman} J.K. Jain, and V.J. Goldman,  Phys. Rev. B {\bf 45}, 1255 (1992).

\bibitem{Quinn}  P. Sitko, K.-S. Yi, and J.J. Quinn, Phys. Rev. B {\bf 56}, 12417 (1997).

\bibitem{jain4over11}  C.-C. Chang, and J.K. Jain, Phys. Rev. Lett. {\bf 92}, 196806 (2004).

\bibitem{lopez2} A. Lopez, and E. Fradkin, Phys. Rev. B {\bf 69}, 155322 (2004).

\bibitem{scomp2} M.O. Goerbig, P. Lederer, and C.M. Smith,  Phys. Rev. Lett.
   {\bf 93}, 216802  (2004).

\bibitem{pashitskii} E.A. Pashitskii, Low Temperature Physics, {\bf 31}, 171 (2005).

\bibitem{rrhaldane}  E.H.  Rezayi, and F.D.M. Haldane,  Phys. Rev. Lett. {\bf 84}, 4685  (2000).

\bibitem{bkdmrg}  E.J. Bergholtz, and A. Karlhede, cond-mat/0304517 (unpublished).

\bibitem{white} S.R. White, Phys. Rev. Lett. {\bf 69}, 2863 (1992); Phys. Rev. B {\bf 48}, 10345 (1993).

\bibitem{kinvariance} B.I. Halperin, and A. Stern, Phys. Rev. Lett. {\bf 80}, 5457 (1998).

\bibitem{taowu} R. Tao, and Y.-S. Wu, Phys. Rev. B {\bf 31}, 6859 (1985).

\bibitem{haldanegap} F.D.M. Haldane,  Phys. Lett. 93A, 464 (1983).

\bibitem{juha} J. Suorsa, private communication.

\bibitem{greiter} M. Greiter, X.G. Wen, and F. Wilczek, Phys. Rev. Lett {\bf 66}, 3205 (1991); Nucl. Phys. B {\bf 374}, 567 (1992).

\bibitem{52exp} R. Willett, J.P. Eisenstein, H.L. St\"ormer, D.C. Tsui, A.C. Gossard, and J.H. English, Phys. Rev. Lett. {\bf 59}, 1776 (1987).

\bibitem{morf} R.H. Morf, Phys. Rev. Lett. {\bf 80}, 1505  (1998).

\bibitem{moller} G. Moller, and S.H. Simon, arXiv:0708.2680v1 (2007).

\bibitem{Freedman}  See {\it eg,} M.H. Freedman, A. Kitaev, M. Larsen, and Z. Wang, Bull. AMS. {\bf 40}, 31 (2003);
A. Kitaev, Ann. Phys. {\bf 303}, 2 (2003); S. Das Sarma, M. Freedman, C. Nayak, S.H. Simon, and A. Stern, arXiv:0707.1889v1 (2007).

\bibitem{apf1} M. Levin, B.I. Halperin, and B. Rosenow, arXiv:0707.0483v1 (2007).

\bibitem{apf2} S.-S. Lee, S. Ryu, C. Nayak, and M.P.A. Fisher, arXiv:0707.0478v1 (2007).

\bibitem{emiljuha}  E.J. Bergholtz, and J. Suorsa, unpublished.

\bibitem{parafermionic} N. Read, and E.H.  Rezayi, Phys. Rev. B {\bf 59}, 8084 (1999).

\bibitem{Haldane85}  F.D.M. Haldane, and E.H. Rezayi, Phys. Rev. B {\bf 31}, R2529 (1985).

\bibitem{fradkin} E. Fradkin,  {\it Field Theories of Condensed Matter Systems}, (Addison-Wesley, 1991).

\bibitem{read} N. Read, and E.H. Rezayi, Phys. Rev. B {\bf 54}, 16864 (1996).


\bibitem{yoshioka} D. Yoshioka, B.I. Halperin, and P.A. Lee, Phys. Rev. Lett. {\bf50}, 1219 (1983).

\bibitem{toruswfs} M. Hermanns, J. Suorsa, E.J. Bergholtz, T.H. Hansson, and A. Karlhede,  Phys. Rev. B {\bf 77}, 125321 (2008).

\bibitem{pfafftorus}  M. Greiter, X.G. Wen, and F. Wilczek,  Nucl. Phys. B {\bf 374}, 567 (1992).

\bibitem{proj} S.M. Girvin, and T. Jach, Phys. Rev. B {\bf 29},  5617 (1984).

\bibitem{thouless} D.J. Thouless, Surf. Sci. {\bf 142}, 147 (1984).

\bibitem{Lieb} E. Lieb, T. Schultz, and D. Mattis, Ann. Phys. {\bf 16}, 407 (1961).

\bibitem{schulz} H.J. Schulz, G. Cuniberti, and P. Pieri,  in {\it Field
Theories for Low-Dimensional Condensed Matter Systems}, eds. G. Morandi et
al.,  (Springer, 2000).


}
\end{thebibliography}
\end{document}